\documentclass[10pt,twoside,twocolumn,english,aps,manuscript,superscriptaddress,prbt,aps,preprint,manuscript,showpacs]{revtex4}
\usepackage{lmodern}
\usepackage{lmodern}
\usepackage[T1]{fontenc}
\usepackage[latin9]{inputenc}
\usepackage{xcolor}
\usepackage{babel}
\usepackage{array}
\usepackage{float}
\usepackage{textcomp}
\usepackage{bm}
\usepackage{amsmath}
\usepackage{graphicx}
\usepackage[unicode=true,pdfusetitle,
 bookmarks=true,bookmarksnumbered=false,bookmarksopen=false,
 breaklinks=false,pdfborder={0 0 1},backref=false,colorlinks=false]
 {hyperref}
\usepackage{breakurl}

\makeatletter

\newcommand*\LyXThinSpace{\,\hspace{0pt}}
\newcommand{\lyxmathsym}[1]{\ifmmode\begingroup\def\b@ld{bold}
  \text{\ifx\math@version\b@ld\bfseries\fi#1}\endgroup\else#1\fi}

\providecommand{\tabularnewline}{\\}

\@ifundefined{textcolor}{}
{%
 \definecolor{BLACK}{gray}{0}
 \definecolor{WHITE}{gray}{1}
 \definecolor{RED}{rgb}{1,0,0}
 \definecolor{GREEN}{rgb}{0,1,0}
 \definecolor{BLUE}{rgb}{0,0,1}
 \definecolor{CYAN}{cmyk}{1,0,0,0}
 \definecolor{MAGENTA}{cmyk}{0,1,0,0}
 \definecolor{YELLOW}{cmyk}{0,0,1,0}
}

\usepackage{babel}
\usepackage{babel}

\makeatother

\begin{document}

\title{Survival of magnetic- correlations above ordering temperature in
a ferromagnetically ordered classical kagomé magnet: Li$_{9}$Cr$_{3}$(P$_{2}$O$_{7}$)$_{3}$(PO$_{4}$)$_{2}$}

\author{R. Kumar }
\email{photon1900@gmail.com}

\affiliation{Department of Physics, Faculty of Science, Hokkaido University, Sapporo
060-0810, Japan}

\author{A. Chakraborty}

\affiliation{Department of Physics, Indian Institute of Technology Kanpur, India}

\affiliation{\textcolor{black}{Institute of Physics, Johannes Gutenberg University,
Mainz, Germany}}

\author{S. Fukuoka}

\affiliation{Department of Physics, Faculty of Science, Hokkaido University, Sapporo
060-0810, Japan}

\author{F. Damay}

\affiliation{Laboratoire Léon Brillouin, Université Paris-Saclay, CNRS, CEA, CE-Saclay,
F-91191 Gif-sur-Yvette, France}

\author{E. Kermarrec}

\affiliation{Université Paris Saclay, CNRS, Laboratoire de Physique des Solides,
91405 Orsay, France}

\author{P. L. Paulose}
\email{paulose@tifr.res.in}

\affiliation{Department of Condensed Matter Physics and Material Sciences, Tata
Institute of Fundamental Research, Colaba, Mumbai - 400005 (India)}

\author{Y. Ihara }
\email{yihara@phys.sci.hokudai.ac.jp}

\affiliation{Department of Physics, Faculty of Science, Hokkaido University, Sapporo
060-0810, Japan}

\date{\today}
\begin{abstract}
Motivated by the recent discovery of a semiclassical nematic spin
liquid state in a Heisenberg kagomé antiferromagnet Li$_{9}$Fe$_{3}$(P$_{2}$O$_{7}$)$_{3}$(PO$_{4}$)$_{2}$
(LFPO) with $S$=5/2 {[}Kermarrec $et$ $al$. Phys. Rev. Lett. \textbf{127},
157202 (2021){]}, we now investigate the impact of spin quantum number
$S$ on the ground state properties by studying the isostructural
kagomé magnet Li$_{9}$Cr$_{3}$(P$_{2}$O$_{7}$)$_{3}$(PO$_{4}$)$_{2}$
(LCPO) with active $t_{2g}$ orbitals and $S$ = 3/2. Thermodynamic
measurements reveal that the ground state properties of LCPO is dominated
by the ferromagnetic interactions with a mean - field temperature
$\varTheta$ $\sim$ 3~K ($J$ < 1~K) and the ordering temperature,
$T$$_{c}$ \textasciitilde{} 2.7~K, and the size of the ordered
moment \textasciitilde ~1.05 $\pm$ 0.25~$\mu_{B}$ is significantly
reduced from that of a fully ordered moment. The \textit{ab initio}
electronic structure calculations nicely corroborate the thermodynamic
results and suggest the presence of additional in and out-of-plane
further neighbor antiferromagnetic couplings, though significantly
weaker in comparison to the dominant first-nearest neighbor ferromagnetic
coupling. The spin-lattice relaxation rate measured with fields larger
than the saturation field shows a magnetic field induced gap ($\Delta$
$\propto$\textit{ B}) in the excitation spectrum, and in $B$$\rightarrow0$
limit the gap has a finite intercept \textasciitilde{} 3~K, equivalent
to the mean-field scale. We interpret the origin of this gap is associated
with the magnetic interactions inherent to the material. With our
experimental results, we establish the stabilization of a ferromagnetic
like ground state and the persistence of magnetic-correlations above
the ordering temperature in LCPO. 
\end{abstract}
\maketitle

\section{Introduction}

The kagomé lattice with antiferromagnetic interactions has the distinction
of being a highly frustrated lattice in two-dimensional (2D) that
defies the Néel order and hosts a highly dynamic state of moments
(spin liquid with emergent excitations). \cite{Balenta1,Emergent excitations 2}
On the other hand, a kagomé lattice with ferromagnetic interactions
does not manifest a classical/quantum spin liquid state, but rather
stabilizes a purely ferromagnetic (FM) ground state. \cite{Heydite-3}
Chisnell \textit{et al}. \cite{Cu(bdc)-4} has demonstrated that the
insulating kagomé ferromagnet has the potential to host topological
properties such as the topological magnon bands as well as a bulk
magnon Hall effect \cite{Cu(bdc)-5 Cp}.

While the low - spin quantum number $S=1/2$ based kagomé materials
have been the primary choice of researchers and were thoroughly studied
in quest of discovering a spin-liquid state \cite{Balenta1,Emergent excitations 2,Heydite-3,Helton Kagome6,Broholom Fractaionalized excitations7,Norman RMP8},
the materials with $S>1/2$ are relatively less explored, but at the
same time are equally interesting and promising to look for some uncharted
phases of quantum magnetism \cite{Hida S1-9,Coupled cluster HAK for S10,HSS ANsatz 11,Spin-S Magnetization12,Spin ordered gs for Cr13,Changlani NaTiO14,Edwin LFPO15}.
For example, be it the idea of a hexagonal-singlet solid state \cite{Hifa HSS16}
or a resonating Affleck-Kennedy-Lieb-Tasaki loop (RAL) \cite{Hao arxiv QQSN 17,TOp=00003D000026 criticality AKLT18}
state in the context of $S=1$ kagomé Heisenberg Antiferromagnet (KHAF),
the appearance of magnetic field induced nematic and supernematic
phases in $S=1$ and $S=2$ kagomé antiferromagnets \cite{NematicS12@19},
a highly debated ground state of a $S=3/2$ kagomé material, SrCr$_{9p}$Ga$_{12-9p}$O$_{19}$,
ranging from a spin glass, spin liquid to a topological glass \cite{Quasi2dSCGO-20,SGCOdilutionGlass21,Kerenmusr-22,Cp SCGO23,SCGOSL 24,LeeGlass SCGO25,GlassSCGO26,SCGO topoglass27},
and the presence of a one-third magnetization plateau in a $S=5/2$
KHAF under higher magnetic fields \cite{Edwin LFPO15}, to name a
few.

We have been investigating a family of kagomé materials with chemical
formula Li$_{9}$M$_{3}$(P$_{2}$O$_{7}$)$_{3}$(PO$_{4}$)$_{2}$
\cite{Poisson CrFe 28} and recently some of us discovered a classical
spin liquid state in the temperature range $T$$_{N}$~(1.3~K) $\leq$
$T$ $\leq$ $\varTheta$~(11~K), and the signature of a one-third
plateau in magnetization of the Fe$^{3+}$~($S=5/2$) based compound
Li$_{9}$Fe$_{3}$(P$_{2}$O$_{7}$)$_{3}$(PO$_{4}$)$_{2}$ (LFPO)
\cite{Edwin LFPO15}. The structure is highly flexible and can accommodate
a variety of magnetic ions M~{[} = V$^{3+}$ ($S=1$), Cr$^{3+}$
($S=3/2$) and Fe$^{3+}$($S=5/2$){]} and nonmagnetic ions M~{[}
= Al$^{3+}$ and Ga$^{3+}${]} while keeping the trigonal symmetry
preserved. The material with quantum number $S=1,$ $i.e.$, Li$_{9}$V$_{3}$(P$_{2}$O$_{7}$)$_{3}$(PO$_{4}$)$_{2}$
is found to exhibit the signature of ferromagnetic fluctuations in
its ground state and this material could be a potential candidate
to stabilize some of the competing magnetic phases by perturbing the
local energy scale \cite{LVPOprep29,LVPOprep30,LVPONMR31,LVPONMR32,LVPONMRTheory33}.
All in all, this family offers a greater tunability to the ground
state properties on changing the orbital selection and the magnitude
of spin quantum number by varying the trivalent ion, so the associated
quantum fluctuations, very much like the extensively studied jarosite
family of kagomé materials, but with an important difference being
in the nature of their ground state properties \cite{V jaro-34,Fe Jaro wills35,musrKerenjaro36,SwFeJaro37,CrjaroESR38}.
More importantly, a low-exchange ($J$ $\leq$ 1~K) makes this family
of materials the suitable candidates to realize some of the magnetic
field induced novel quantum phases.

Herein, we investigate the structural and thermodynamic properties
of a \textit{structurally clean} two - dimensional kagomé material
Li$_{9}$Cr$_{3}$(P$_{2}$O$_{7}$)$_{3}$(PO$_{4}$)$_{2}$ (LCPO)
\cite{Poisson CrFe 28} by employing a number of probes including
x - ray, neutron diffraction, magnetization, specific heat and NMR.
To our surprise, the kagomé network decorated with Cr$^{3+}$ ions
($S=3/2$) stabilizes a ferromagnetic - like ground state. In this
study, we will highlight the possible reasons behind the stability
of a ferromagnetic order in LCPO.

\section{Experimental and theoretical details }

The polycrystalline samples of LCPO were synthesized by mixing and
heating the materials: LiH$_{2}$PO$_{4}$ (Alfa Aeasar: 97\% purity),
Cr$_{2}$O$_{3}$(Alfa Aeasar: 99.98\% purity) and Li$_{4}$P$_{2}$O$_{7}$
in stoichiometry at different temperatures for varying duration (in
hours): 250${^\circ}$C\,(6~h), 500${^\circ}$C\,(12~h), 700${^\circ}$C\,(12~h),
730${^\circ}$C\,(12~h) and 750${^\circ}$C\,(24~h) on a platinum
foil in a programmed Carbolite furnace. The Li$_{4}$P$_{2}$O$_{7}$
was prepared by baking a stoichiometric amount of Li$_{3}$PO$_{4}$
(Sigma-Aldrich: >~99.9\% purity) and (NH$_{4}$)H$_{2}$PO$_{4}$
(Sigma-Aldrich: >~99.99\% purity) in the temperature range 300 -
605${^\circ}$C\,(24h) with several intermittent grindings. The phase
purity and the structural determination was carried out using the
data collected on a PANalytical X'Pert PRO diffractometer at room
temperature using Cu K$_{\alpha}$ radiation ($\lambda$ = 1.5418\,$\textrm{Å}$)
and also on the neutron diffraction data (acquired on the G4.1 diffractometer
at the LLB) collected in the temperature range 1.65 - 80~K with a
wavelength $\lambda=2.426$\,$\textrm{Å}$. The dc magnetization
measurements were carried out on a Quantum Design MPMS XL system and
the heat capacity was measured using thermal - relaxation method in
the temperature range 1.8 - 230~K utilizing the heat capacity option
of a Quantum Design PPMS, whereas in the temperature range 0.6 - 10~K,
a home - made calorimeter was used, and the low temperature $C$$_{p}$
data were scaled in the common temperature range. The ac susceptibility
measurements under pressure (0 - 9~kbar) were done on a Quantum Design
MPMS XL system using the pressure cell option from the $easylab$.
NMR measurements were performed using a home-made spectrometer constructed
with the software-defined radio technology \cite{Spectrometer39}.
For the field-sweep NMR spectrum measurements, we used a 14 T-Oxford
magnet energized by IPS120 power supply, which can sweep the magnetic
fields at a constant rate.

\textcolor{black}{The electronic structure calculations based on density
functional theory (DFT) presented in this paper are carried out in
the plane-wave basis within generalized gradient approximation (GGA)
\cite{ref40} of the Perdew-Burke-Ernzerhof exchange correlation supplemented
with Hubbard U as encoded in the Vienna ab-initio simulation package
(VASP) \cite{ref41,ref42} with projector augmented wave potentials
\cite{ref43,ref44}. The LCPO crystal consists of two formula units
with a total of 98 atoms within the unit cell. The calculations are
done with usual values of U and Hund's coupling ($J_{H}$) chosen
for Cr with $U_{eff}$ ($\equiv U-J_{H}$) $=3.5$ eV in the Dudarev
scheme \cite{ref45}. In order to achieve convergence of energy eigenvalues,
the kinetic energy cut off of the plane wave basis was chosen to be
$500$~eV. The Brillouin-Zone integration is performed with $4\times4\times4$
Monkhorst grid of k-points.}

\section{results and discussion}

\begin{figure}
\begin{centering}
\includegraphics[scale=0.34]{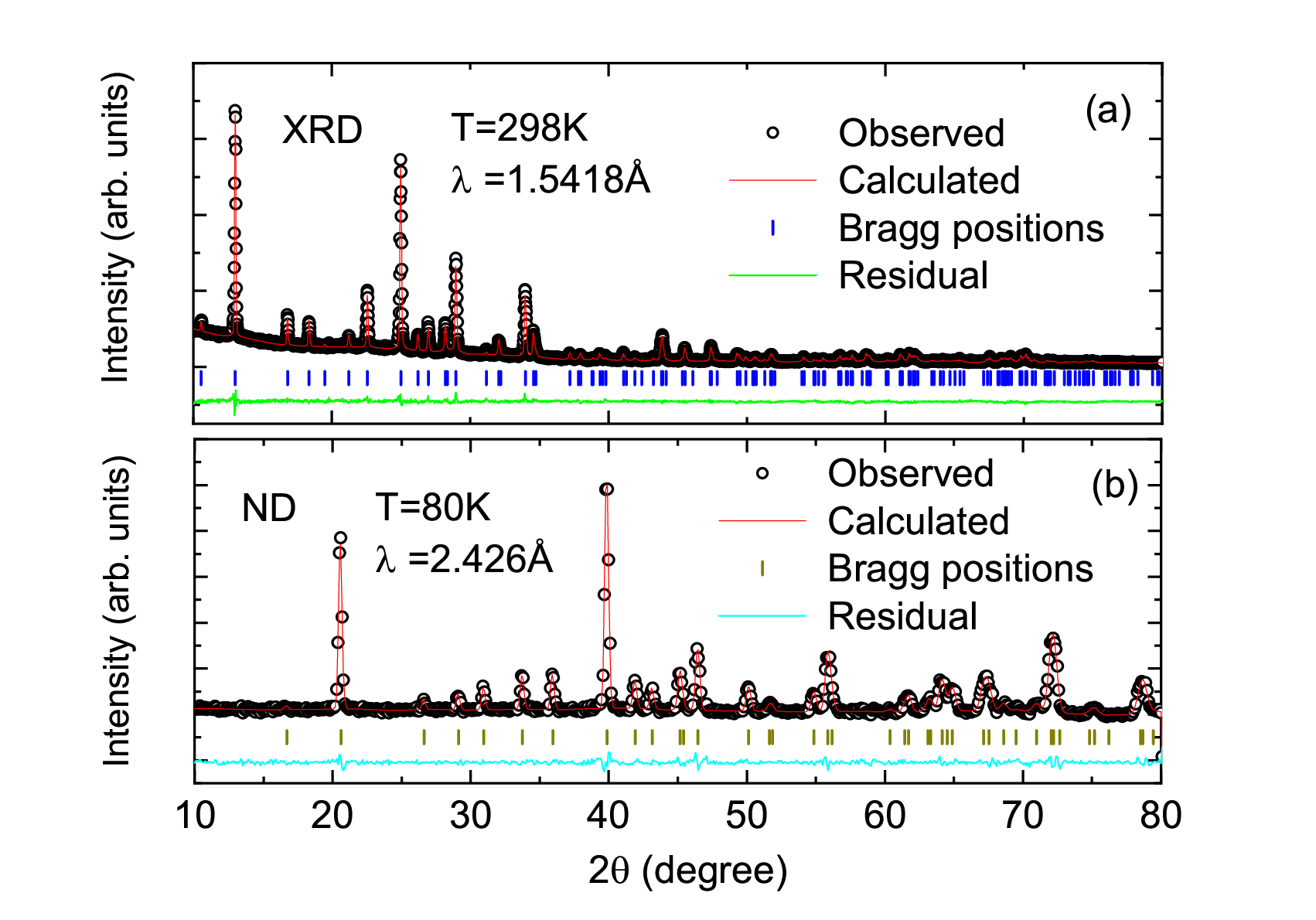} 
\par\end{centering}
\caption{\label{XRD=00003D000026ND} (a) Powder x-ray diffraction profile of
LCPO collected at $T$ $=29$8\,K with $\lambda$=1.5418~$\textrm{\textrm{Å}}$.
(b) Neutron diffraction profile collected at $T$ $=$80\,K with
$\lambda$=2.426~$\textrm{Å}$.}
\end{figure}

X-ray diffraction data collected on a polycrystalline sample of LCPO
is shown in Fig. \ref{XRD=00003D000026ND}~(a). The data were refined
using FullProf Suite program \cite{FP-46} under the centrosymmetric
space group $P\bar{3}c1$ (no. 165) and the extracted lattice parameters
$a$ = $b$ = 9.6654(2)~$\lyxmathsym{\AA}$ and $c$ = 13.5904(1)~$\lyxmathsym{\AA}$
are in an excellent agreement with the literature reported values
\cite{Poisson CrFe 28}. The atomic position refinement using x-ray
diffraction data, however, lead to the over-refined values of oxygen
atomic positions, violating the Shannon ionic radii constraint \cite{Shannon47}
for P(1) - O(2) \& P(2) - O(4) bonds. We then performed neutron diffraction
experiments to reliably estimate the atomic positions and the neutron
diffraction pattern obtained at 80~K is shown in Fig. \ref{XRD=00003D000026ND}~(b)
and the refinement results are summarized in Table \ref{Li9Cr3P8O29 unit cell parameters-ND}.
The parameters defining the quality of experimental fit are obtained
to be $R$$_{p}$ = 2.35\%, $R$$_{wp}$ = 3.05\%, $R$$_{exp}$ =
2.24\%, and $\chi^{2}$ = 1.85.

\begin{table*}
\begin{centering}
\caption{\label{Li9Cr3P8O29 unit cell parameters-ND} Unit-cell parameters
for LCPO after neutron diffraction refinement at $T$ = 80~K with
($\lambda$= 2.426\,$\textrm{Å}$) and the obtained lattice constants
are $a$=$b$= 9.6492(8)~$\textrm{\textrm{Å}}$ and $c$= 13.5654(2)~$\textrm{\textrm{Å}}$. }
\par\end{centering}
\medskip{}

\centering{}%
\begin{tabular}{l>{\centering}p{1cm}>{\centering}p{1.8cm}>{\centering}p{1.8cm}>{\centering}p{1.8cm}>{\centering}p{0.85cm}>{\centering}p{0.8cm}}
\hline 
Atoms  & site  & \textcolor{black}{x}  & y  & z  & Biso  & Occupancy\tabularnewline
\hline 
P(1)  & 4d  & 2/3 & 1/3 & 0.6279(32)  & 0.53  & 1\tabularnewline
P(2)  & 12g  & 0.3213(23)  & 0.0893(27)  & 0.8409(20)  & 0.73  & 1\tabularnewline
Cr  & 6f  & 0.5780(61)  & 0.0  & 0.75  & 0.22  & 1\tabularnewline
Li(1)  & 2b  & 0.0  & 0.0  & 0.0  & 1.97  & 1\tabularnewline
Li(2)  & 4d  & 2/3 & 1/3 & 0.8639(76)  & 0.71  & 1\tabularnewline
Li(3)  & 12g  & 0.3524(63)  & 0.1009(64)  & 0.0751(54)  & 0.91  & 1\tabularnewline
O(1)  & 4d  & 2/3 & 1/3 & 0.5194(26)  & 0.71  & 1\tabularnewline
O(2)  & 6f  & 0.2039(34)  & 0.0  & 0.75  & 0.77  & 1\tabularnewline
O(3)  & 12g  & 0.6752(18)  & 0.1827(20)  & 0.6618(18)  & 0.63  & 1\tabularnewline
O(4)  & 12g  & 0.4851(29)  & 0.1088(22)  & 0.8279(20)  & 0.71  & 1\tabularnewline
O(5)  & 12g  & 0.3308(15)  & 0.2546(26)  & 0.8464(14)  & 0.27  & 1\tabularnewline
O(6)  & 12g  & 0.2309(22)  & -0.0064(20)  & 0.9310(15)  & 0.67  & 1\tabularnewline
\hline 
\end{tabular}
\end{table*}

\begin{figure}
\centering{}\includegraphics[scale=0.17]{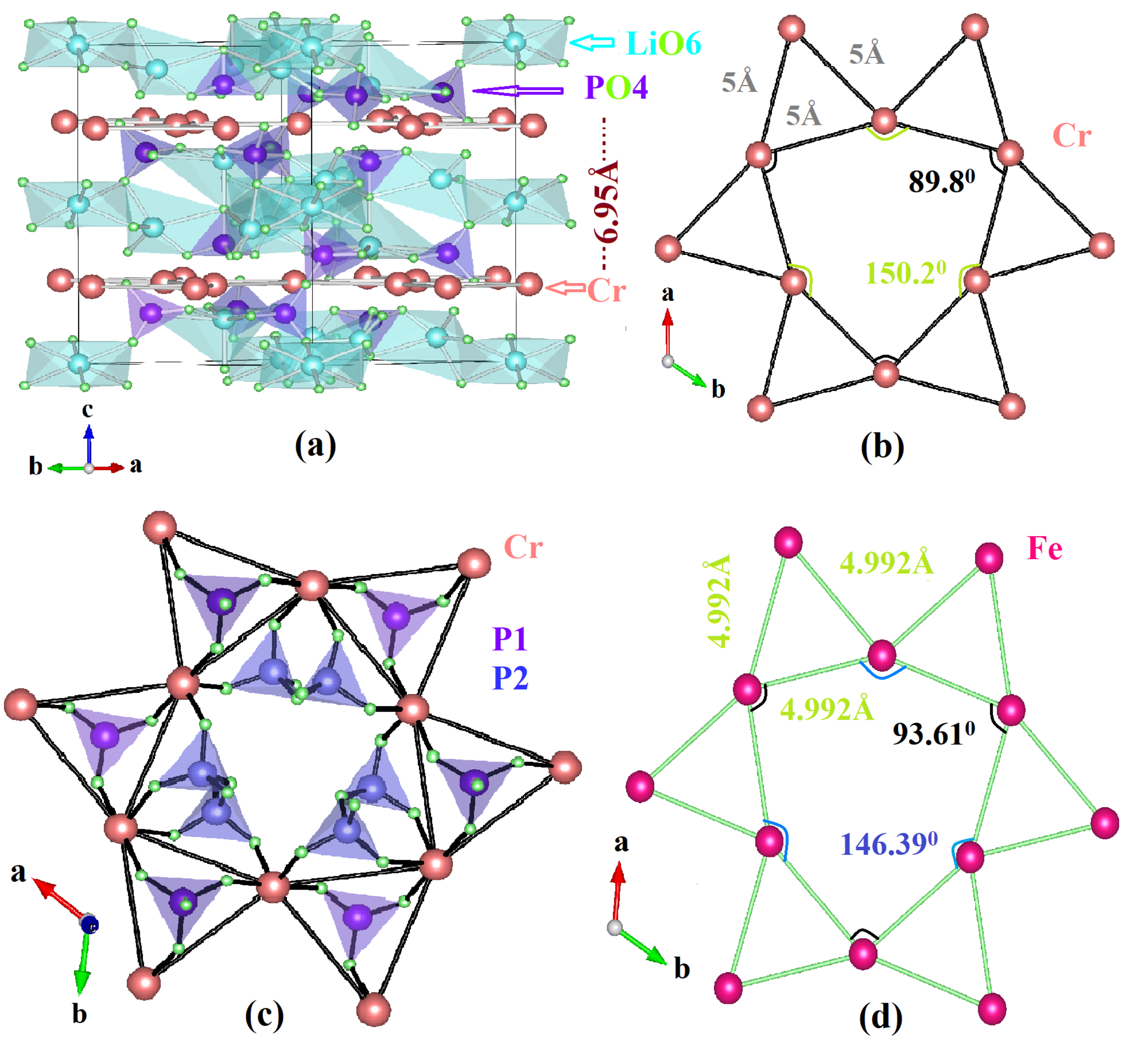}\caption{\label{Structure} The crystal structure of LCPO drawn using VESTA
program \cite{Vesta48}. (a) A representative unit-cell of LCPO with
kagomé layers (Cr) separated by LiO$_{6}$ octahedra (light cyan)
and PO$_{4}$ tetrahedra (light violet). (b) A view of kagomé plane
formed by Cr - atoms in $ab$ - plane. (c) Interaction pathways between
Cr - atoms, for a kagomé motif, mediated through PO$_{4}$ bridges.
(d) A view of kagomé plane formed by Fe - atoms (in LFPO) in $ab$
- plane. (Note: P1~(violet) is located apically the center of each
triangle and its position changes alternatively from $z$ to $-z$
on moving from one to the next triangle, while P2~(light blue) is
structurally positioned in the kagomé hexagon)}
\end{figure}

Structural Details: Edge sharing CrO$_{6}$ octahedra form kagomé
network in $ab-$ plane and in a unit cell of LCPO consecutive kagomé
planes are \textasciitilde{} 6.95~$\textrm{Å}$ apart and separated
by nonmagnetic LiO$_{6}$ and PO$_{4}$ units, see Fig. \ref{Structure}~(a).
The kagomé lattice formed by the equilateral triangles (side $\sim$
5$\,\textrm{Å}$: \textcolor{black}{forming the nearest neighbor (nn)
connectivity}) has a hugely deformed kagomé hexagon with the lowest
and highest angles being 89.8$^{{^\circ}}$ \& 150.2$^{{^\circ}}$,
see Fig. \ref{Structure}~(b), while for a regular kagomé pattern,
vertices in a hexagon tend to make an angle of 120.0$^{{^\circ}}$.
Despite the kagome deformation, the structure possesses a unique magnetic
site with equilateral triangles preserving the kagome physics. The
magnetic interaction between Cr - ions is mediated via extended pathways
(Cr - O - P - O - Cr) provided by the PO$_{4}$ bridges, see Fig.
\ref{Structure}~(c). In particular, two crystallographically inequivalent
phosphorus atoms --- P(1) at (4g) and P(2) at (12d) --- are responsible
for facilitating the interactions between intra - triangle and kagomé
network. Magnetic interaction is expected to be weak in view of the
extended pathways. On comparing the LCPO with LFPO, it is noticeable
that the distance between Fe - ions forming the kagomé network is
slightly reduced $\sim$ 4.992$\lyxmathsym{\AA}$ and also the kagomé
hexagon is marginally less distorted, Fig. \ref{Structure}~(d).
Notably, the LCPO has a reduced unit cell ($a$ = $b$ = 9.6654(2)~$\lyxmathsym{\AA}$,
$c$ = 13.5904(1)~$\lyxmathsym{\AA}$, and V = 1099.536(24)~$\lyxmathsym{\AA}^{3}$)
with respect to the LFPO, which has a slightly larger unit - cell
($a$ = $b$ = 9.7195(8)~$\lyxmathsym{\AA}$, $c$ = 13.6059(2)~$\lyxmathsym{\AA}$)
with cell volume V = 1113.128(22)~$\lyxmathsym{\AA}^{3}$.

\subsection{Magnetization}

\begin{figure}
\begin{centering}
\textcolor{black}{\includegraphics[scale=0.33]{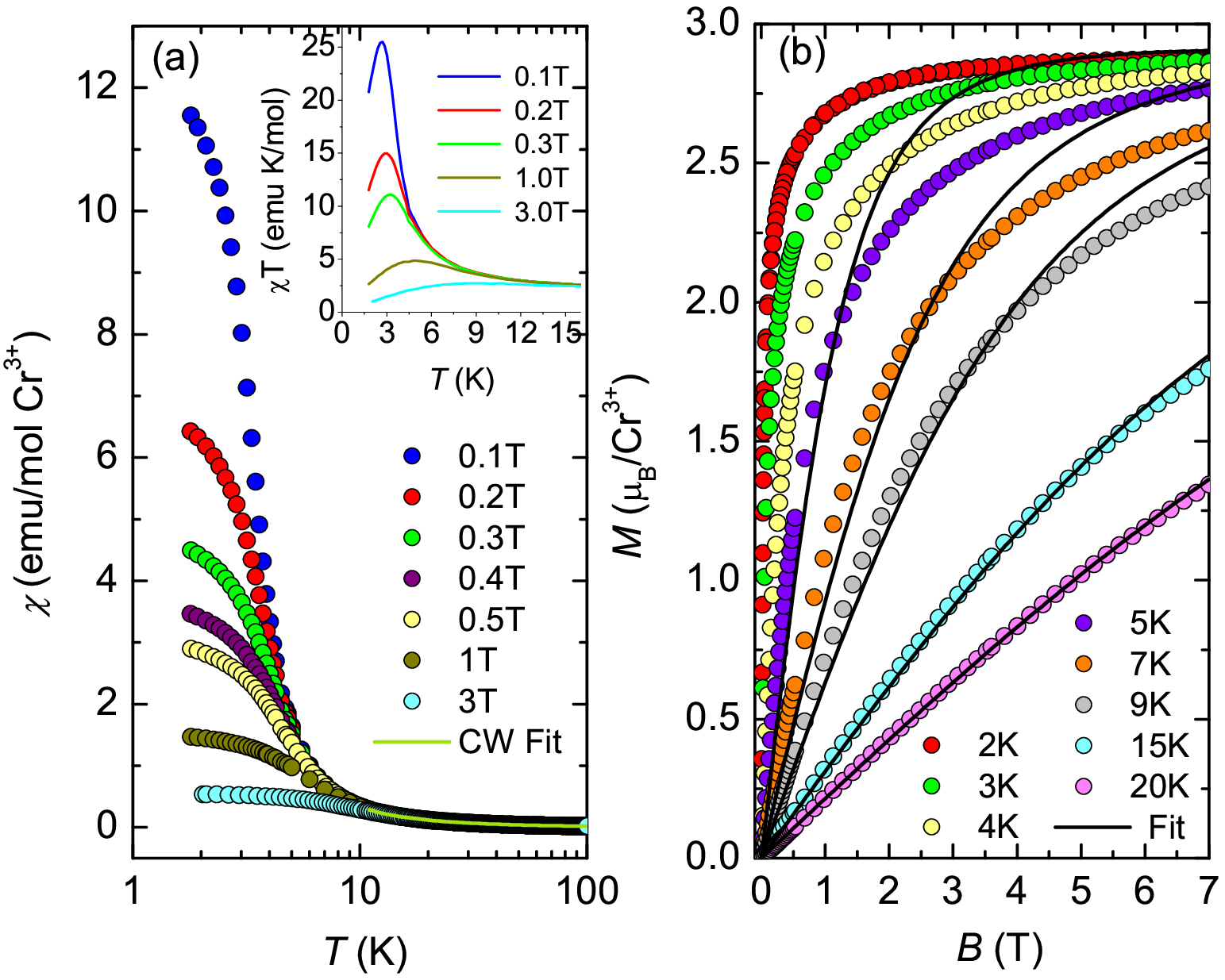} }
\par\end{centering}
\textcolor{black}{\caption{\label{Magnetization} (a) dc susceptibility as a function of temperature
in the field range 0.1 - 3~T and the solid line (green) is a fit
to the Curie-Weiss law. Inset: $\chi$$T$ vs $T$ plot. (b) Magnetic
isotherms (filled circle) and their fit to the Brillouin function
(black solid line) in the field range 0 - 7~T.}
}
\end{figure}

The temperature and field dependence of dc magnetic susceptibility
($\chi=M/H$) measured in the temperature range 1.8 - 300~K and field
range 0.1 - 3~T is shown in Fig. \ref{Magnetization}~(a). $\chi(T)$
shows a weak temperature dependence in a large temperature region
(10 - 100~K) and then abruptly increases below \textasciitilde{}
10~K, indicating a possible thermodynamic phase transition. A fit
to the Curie-Weiss law, $\chi$ = $\chi_{0}$ + $C/(T-\varTheta$),
in the temperature range 10 - 300~K \cite{Curie-Fit49} results in
a temperature independent susceptibility ($\chi_{0}$= (-6.6158 $\pm$
0.61)$\times$10$^{-4}$~emu/mol Cr), Curie constant ($C$ = 1.839(4)~emuK/mol
Cr), and Curie-Weiss temperature ($\varTheta$ = 4.54(2)~K). A positive
value of $\varTheta$ is indicative of the fact that the predominant
interaction is ferromagnetic in nature and the estimated value of
paramagnetic moment, $\mu_{{\rm eff}}$ = 3.84 $\pm$ 0.18~$\mu_{B}$,
is in close agreement with the spin-only moment value, $\mu_{{\rm eff}}$
= 3.87~$\mu_{B}$, as expected for $S=3/2$.\textcolor{black}{{} The
role of spin-orbit coupling is expected to be much weaker for half-filled
Cr$^{3+}$ ($t_{2g}^{3}$e$_{g}^{0}$: $S=3/2$; high-spin electronic
configuration) and so was the case with sister material, LFPO, which
has partially filled Fe$^{3+}$($t$$^{3}$$_{2g}$$e$$^{2}$$_{g}$:
$S=5/2$) orbitals. Interestingly, despite having a compact unit-cell,
the LCPO shows a reduced }Curie-Weiss temperature\textcolor{black}{{}
in comparison to LFPO (}$\varTheta$ = -11~K\textcolor{black}{),
inferring the importance of markedly different exchange pathways and
relative angles governing the magnetic properties. Magnetic isotherms
measured in the temperature range 2 - 20~K (Fig. \ref{Magnetization}~(b))
do not exhibit the signature of any hysteresis and the material gets
nearly fully polarized ($g$$\mu_{B}$$S$ = 3~$\mu_{B}$) with the
saturation moment $M_{S}$ =} 2.91~$\mu_{B}$\textcolor{black}{{} \&
$g$ $\sim$ 1.94 on applying the magnetic field of strength a little
over 1~T at 2~K, possibly hinting either ($i$) the soft nature
of ferromagnetism or ($ii$) the presence of antiferromagnetic coupling
in addition to the dominant nearest - neighbor ferromagnetic coupling.}
A rounded peak in the $\chi T$ vs $T$ plot (inset of Fig. \ref{Magnetization}~(a))
for fields $B$ $\leq$ 1~T is a manifestation of the simultaneous
presence of both the ferro and antiferromagnetic interactions, which
on increasing the field value gradually diminishes in magnitude and
smoothly connects to the paramagnetic region. It is to be noted that
for fields $B$ $\geq$1~T, the ground state gets completely polarized
and the material as a whole essentially starts behaving like a polarized
ferromagnet. The presence of additional couplings has nicely been
captured in our electronic structure calculations (see section D)
and the obtained results are in conjunction with the magnetization
data.\textcolor{black}{{} }
\begin{table}[b]
\textcolor{black}{\caption{\textcolor{black}{\label{Brillouin Fit}} \textcolor{black}{Fitting
results of magnetic isotherm data with the Brillouin function for
$g$ =1.94.}}
}
\centering{}\textcolor{black}{}%
\begin{tabular}{|c|c|c|c|c|c|}
\hline 
\textcolor{black}{Isotherm data (K)} & \textcolor{black}{20} & \textcolor{black}{15} & \textcolor{black}{9} & \textcolor{black}{7} & \textcolor{black}{5}\tabularnewline
\hline 
\hline 
\textcolor{black}{{} $\theta$(K)} & \textcolor{black}{5.36(2)} & \textcolor{black}{4.97(3)} & \textcolor{black}{4.11(6)} & \textcolor{black}{3.74(6)} & \textcolor{black}{3.43(6)}\tabularnewline
\hline 
\end{tabular}
\end{table}

Next, we fitted the $M$($H,$$T$) data to the following equation:
$M(H,T)=$ $N_{A}\mu_{B}SgB_{S}(gSH\mu_{B}/k_{B}(T-\theta))$, where
the symbols $N_{A}$, $\mu_{B}$, $S$, $g$, $B_{S}$, $k_{B}$,
and\textcolor{black}{{} }$\theta$ are the Avogadro number, the Bohr
magneton, spin, the spectroscopic splitting factor, the Brillouin
function, the Boltzmann constant, and correlation temperature, respectively.
\textcolor{black}{The $\theta$ was the only free parameter in the
fit and the fitting values are tabulated in Table \ref{Brillouin Fit}.
The fitting of the magnetic isotherms results in an average value
of $\theta$$\sim$ 4.5$\pm$0.75~K. It is evident from Fig. \ref{Magnetization}~(b)
that the fitting starts deviating below 15~K, possibly highlighting
the survival of magnetic correlations even above the mean-field temperature
$\varTheta$ = 4.54(2)~K perhaps due to competing interactions. }

\textcolor{black}{}

\begin{figure}
\begin{centering}
\includegraphics[scale=0.34]{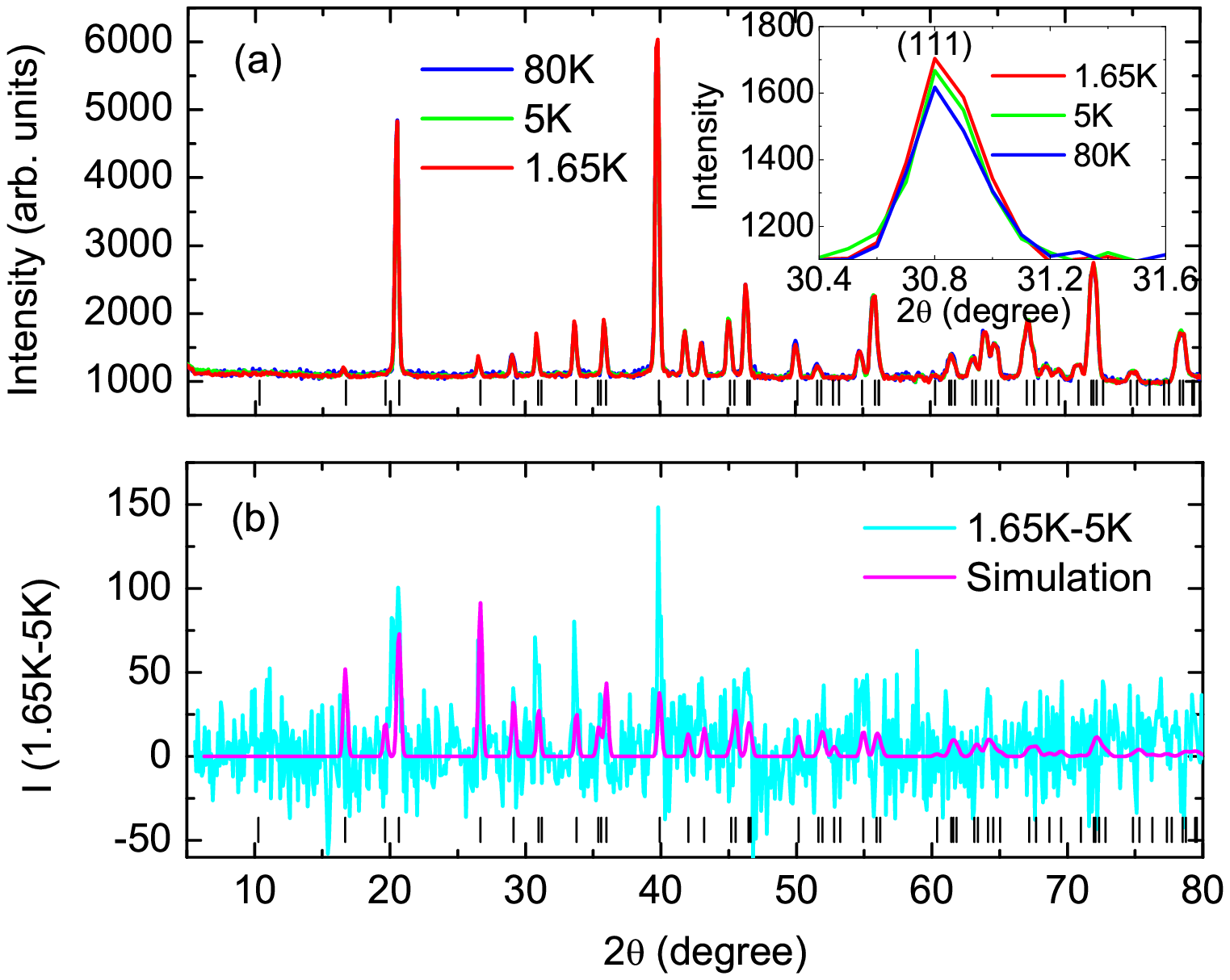} 
\par\end{centering}
\caption{\label{ND and ordered Cr moment} (a) Neutron diffraction profiles
measured at $T$ = 80, 5, and 1.65~K.\textcolor{blue}{{} }\textcolor{black}{Inset
depicts the zoomed portion of diffraction profile for the (111) Bragg
peak.} (b) Difference profile (cyan line) after subtracting out the
5~K data from the 1.65~K data. The pink solid line represents a
simulated pattern for a ferromagnetic model with $k$ = 0 and ordered
moment 1.05~$\mu_{B}$. The negative intensity in the vicinity of
some of the Bragg positions results due to a noisy background. }
\end{figure}

\subsection{Neutron diffraction}

\begin{figure}
\begin{centering}
\includegraphics[scale=0.07]{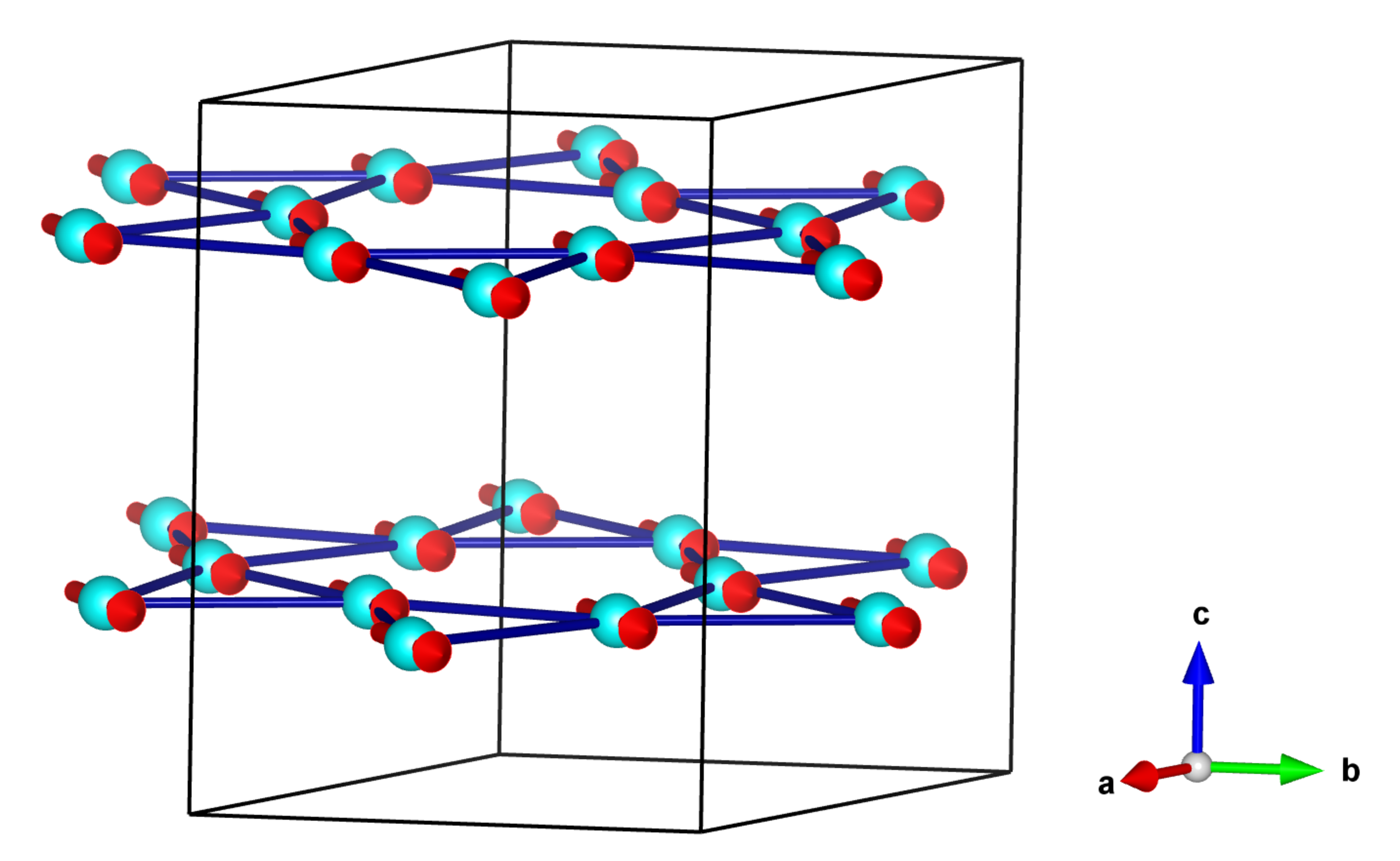}
\par\end{centering}
\caption{\label{magnetic structure} A possible magnetic structure for LCPO
with the resultant moment lying in the $ab$-plane.}

\end{figure}

The neutron diffraction profiles were collected down to 1.65~K, shown
in Fig. \ref{ND and ordered Cr moment}~(a). The absence of extra
reflections other than the nuclear Bragg peaks below $T$$_{c}$ signifies
the absence of an antiferromagnetic ordering. However, a subtle enhancement
in the intensities of a fewer reflections was observed at 1.65~K
compared to 5~K and 80~K and the difference profile, Fig. \ref{ND and ordered Cr moment}~(b),
makes this non-negligible increment more evident. A relatively small
enhancement in the intensity of some of the Bragg peaks is compatible
with a ferromagnetic model with the propagation vector $k$ = 0. The
representational analysis using the BasIreps option of FullProf program
reveals six irreducible representations (IRrep) for the centrosymmetric
trigonal space group ($P\bar{3}c1$). Of which, IRrep(5) and IRrep(6)
found to generate reflections at most of the Bragg positions, but
IRrep(5) was discarded as it missed out one of the most intense reflections
at \textasciitilde{} 20.5$^{\lyxmathsym{\textdegree}}$. The simulated
pattern with IRrep(6) is shown in Fig. \ref{ND and ordered Cr moment}~(b)
and the resultant magnetic structure is shown in Fig. \ref{magnetic structure}.\textcolor{blue}{{}
}\textcolor{black}{The data statistics is neither sufficient enough
to reliably extract the absolute size of the ordered moment ($\mu_{{\rm ord}}$)
nor to discard a moderate spin canting, but our best estimate gives
a value $\sim$~1.05 $\pm$ 0.25~$\mu_{B}$ lying in the $ab$-plane.}
It is highly likely that the missing moment could be distributed in
the inelastic or quasi-elastic channel of scattering in the form of
diffuse scattering. 

\subsection{Heat Capacity}

\begin{figure}
\begin{centering}
\includegraphics[scale=0.35]{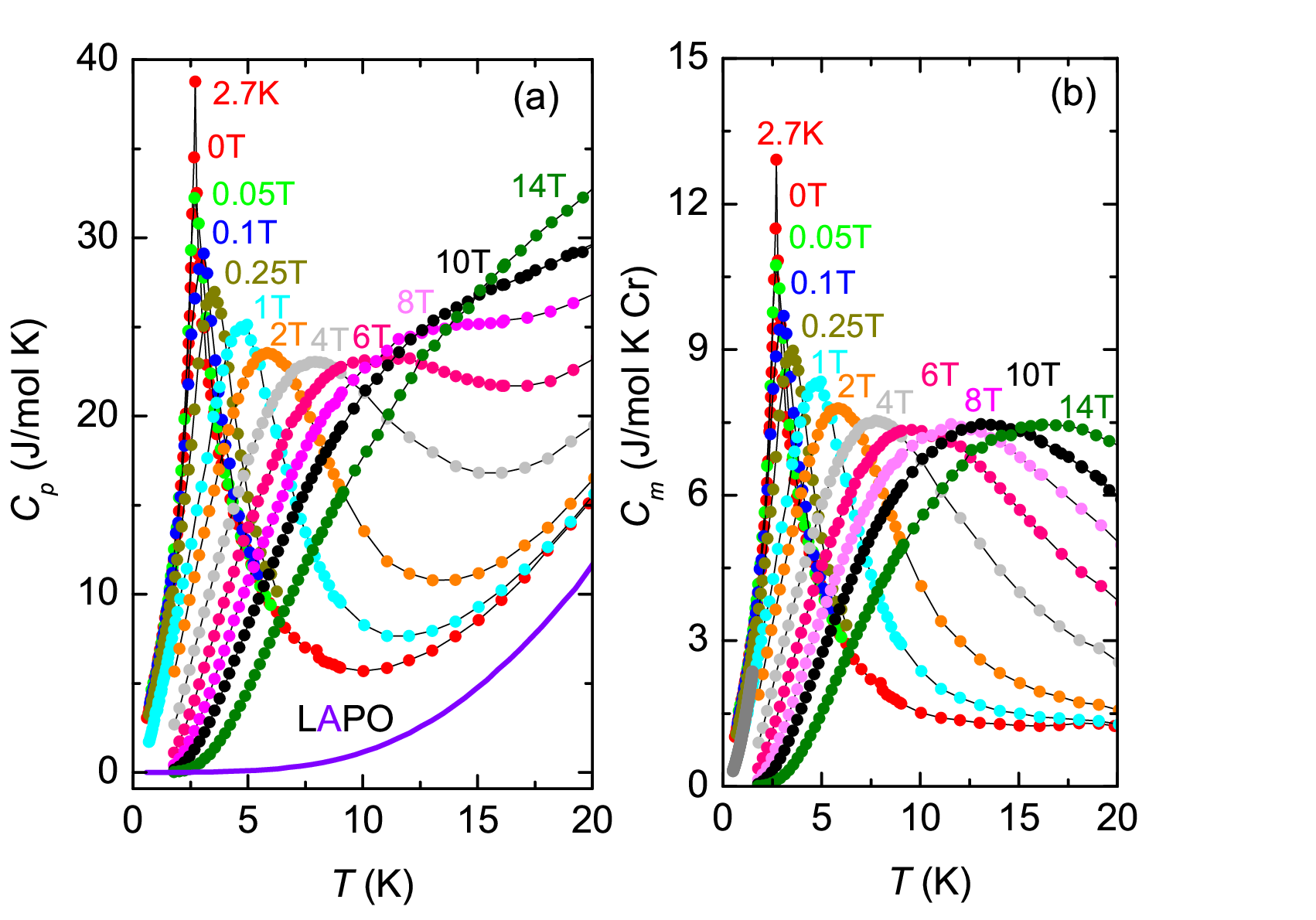} 
\par\end{centering}
\caption{\label{Heat capacity} (a) $C$$_{p}$ versus $T$ in the magnetic
field range 0 - 14~T for LCPO and for nonmagnetic sample LAPO at
$B$ = 0~T. (b) Magnetic specific heat, C$_{m}$, after subtracting
out the phononic contribution. }
\end{figure}

\begin{figure}
\centering{}\includegraphics[scale=0.35]{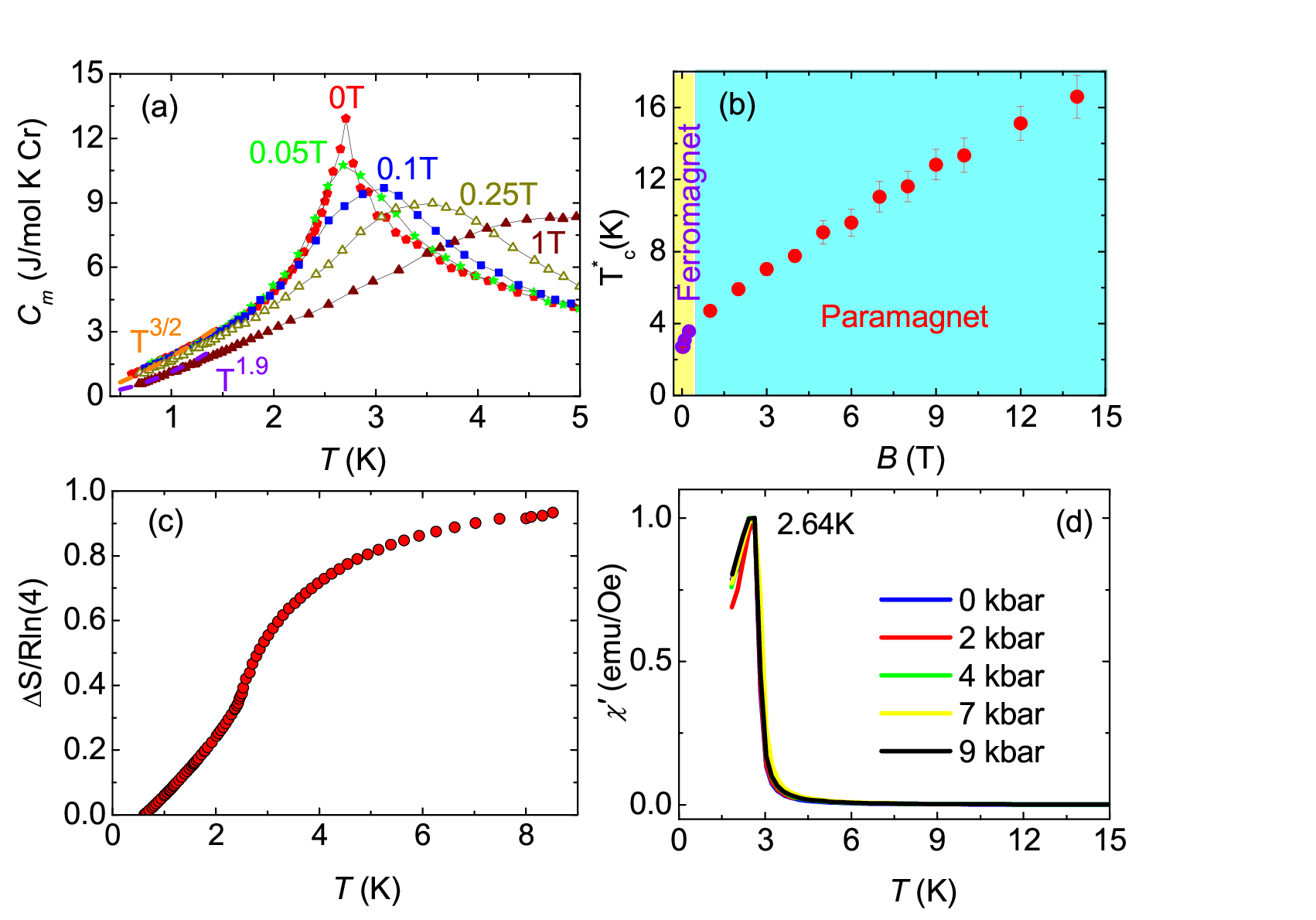}\caption{\label{Specific heat1} (a) $C_{m}$ and its fit to the equation:
$C$$_{m}$ $\propto$ $T$$^{\alpha}$ for fields 0 $\protect\leq$
$B$ $\protect\leq$1~T; orange and violet dashed lines denote fit
to the $T$$^{\frac{3}{2}}$ and $T$$^{1.9}$, respectively. (b)
Field-induced shift in the peak position of anomaly defined by the
temperature ($T_{c}$ = $B$ $\protect\leq$ $B$$_{c}$ \& $T_{c}^{*}$
= $B$ $\protect\geq$ $B$$_{c}$). (c) Entropy change ($\triangle S$)
at $B$ = 0~T in the $T$ - range 0.6 - 10~K. (d) Real component
of ac - susceptibility in the pressure range 0 - 9~kbar. ($B$$_{c}$:
the critical value of magnetic field required to perturb the ground
state).}
\end{figure}

In an attempt to better understand the ground state and the nature
of its excitations, we performed specific heat measurements in the
temperature-range 0.6 - 230~K and the field-range 0 - 14~T and the
results are shown in Fig. \ref{Heat capacity}~(a). A sharp anomaly,
hallmark of a thermodynamic phase transition, was seen at 2.7~K in
$C_{p}$ data on measuring with zero external field. However, the
anomaly was found to get broadened and gradually shifted towards higher
temperature side upon stepping up the external field. The magnetic
specific heat, $C$$_{m}$~( $=C$$_{p}$ - $C_{{\rm lattice}}$),
was extracted after subtracting out the lattice specific heat contribution
from the total specific heat, $C_{p}$. An iso-structural nonmagnetic
material Li$_{9}$Al$_{3}$(P$_{2}$O$_{7}$)$_{3}$(PO$_{4}$)$_{2}$
(LAPO) was measured for estimating the $C_{{\rm lattice}}$ and its
Debye temperature was renormalized in accordance with the magnetic
material by employing the Bouvier scaling \cite{Cpscaling50}. As
shown in Fig. \ref{Heat capacity}~(b) and Fig. \ref{Specific heat1}~(a),
the transition temperature, $T$$_{c}$ ( \textasciitilde{} 2.7~K),
remains field invariant in the narrow field range 0 - 0.05~T, but
the anomaly shifts towards higher temperatures and also gets broadened
on increasing the magnetic field from 0 to 14~T. Such a field-induced
elevation of the transition temperature is akin to the Schottky -
behavior and is a typical signature for a ferromagnetic system, as
for an antiferromagnet the transition temperature gets pushed down
to the lower temperatures on increasing the field strength. A gradual
reduction in the Schottky's peak amplitude with the applied magnetic
field until $B$~$\geq$ ~8~T is also apparent from Fig. \ref{Heat capacity}~(b).
The field dependent variation of the peak-shift is shown in Fig. \ref{Specific heat1}~(b)
and it is apparent that for fields ($B$ $\geq$ 0.25~T) the peak
temperature, labeled as $T_{c}^{*}$, linearly scales with the external
field strength, suggesting that the Zeeman energy scale becomes comparable
to the exchange-coupling ($J$) for $B$ $\geq$ 0.25~T, thus above
this field the ground state essentially starts deviating from a true
ferromagnet and transforms into a partially polarized state. In view
of the observed magnetic ordering temperature (2.7~K), we also calculated
the strength of magnetic dipole interaction term (\textasciitilde{}
$\mu^{2}$/$4\pi r^{3}$) and the obtained value of dipolar energy
scale ($E$$_{dip}$/$k$$_{B}$ \textasciitilde{} 0.04~K) was found
to be two orders of magnitude lower than the ordering temperature,
which clearly means that the magnetic ordering is mainly dominated
by the exchange interactions.

In the framework of spin wave theory, for a dispersion of the type:
$\omega\,$$\propto\,$$k$$^{\phi}$, the magnetic specific heat
is shown to have a temperature dependence of the form: $C$$_{m}$
$\propto$ $T$$^{\alpha}$; $\alpha=\frac{d}{\phi}$, where $d$
and $\phi$ are the magnetic dimensionality of lattice and exponent
in dispersion relation, respectively. The $C_{m}$ was found to have
a  $T$$^{1.3}$ dependence (not shown) in the temperature range 1.5-0.6~K
at $B=0$~T and retains the same dependence in the field range 0
- 0.1~T. The obtained temperature dependence of $C_{m}$ is very
close to the expected \textasciitilde{} $T$$^{\frac{3}{2}}$ behavior
(anticipated for a 3d ferromagnet) \cite{FMCp42}, see Fig. \ref{Specific heat1}~(a).
According to the Mermin-Wagner theorem \cite{Mermin Wagner52}, a
long-range magnetic order (LRO) can not be realized in the 2D isotropic
Heisenberg model at $T$ > 0~K because of the long-range magnetic
fluctuations, and therefore a finite 3D coupling between the kagome
layers is required to uphold a magnetic transition at $T$ > 0~K.
On the other hand, for $B$ = 1~T, $C_{m}$ varies as \textasciitilde{}
$T$$^{1.9}$, which reflects the field induced perturbation in the
excitation spectrum. The magnetic entropy change estimated in the
temperature range 0.6 and 10~K yields \textasciitilde{} 95\% of $R$ln(4),
expected for $S$ = 3/2, see Fig. \ref{Specific heat1}~(c). In fact,
the recovery of $\gtrsim$ 35\% entropy above the mean-field temperature
commensurate with the magnetic-isotherm analysis and further signifies
the persistence of magnetic correlations. The survival of magnetic
correlations above $\varTheta$ was also seen in a $S=3/2$ quasi-2D
honeycomb material Ag$_{3}$LiMn$_{2}$O$_{6}$ \cite{Ag3LiMn2O643}. 

ac-susceptibility measurements were performed under pressure, with
an intention of exploring the possibility of a pressure induced criticality.
The real component of ac susceptibility measured ($H$$_{ac}$= 1~Oe,
$f$ = 11~Hz, $H$$_{dc}$ = 0) in the pressure range 0 - 9~kbar
is shown in Fig. \ref{Specific heat1}~(d). However, no appreciable
change in the transition temperature \textasciitilde 2.64(10)~K
was observed and the transition was found to be robust in the measured
pressure range.

\subsection{Electronic structure calculation}

\begin{figure}[h!]
\textcolor{black}{\centering \includegraphics[width=1\linewidth]{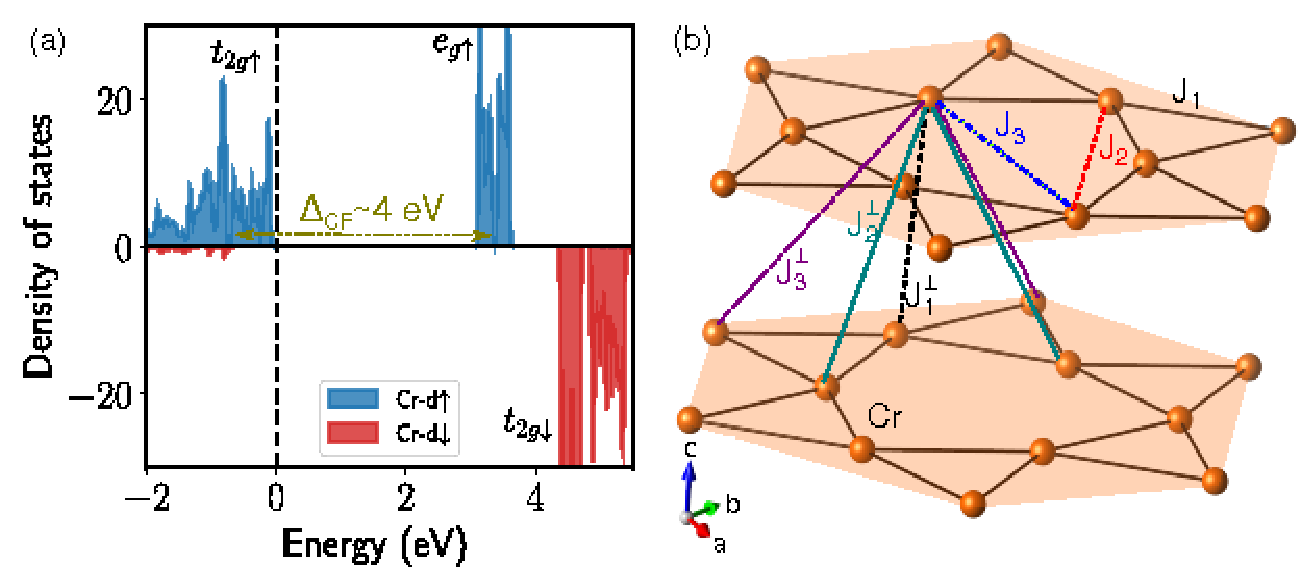}
\caption{(a) Orbital resolved spin-polarized density of states plot of Cr-$d$
orbitals within GGA+U calculations. (b) First, second and third neighbor
intra-planar ($J_{1}$, $J_{2}$ and $J_{3}$) and inter-planar ($J_{1}^{\perp}$,
$J_{2}^{\perp}$ and $J_{3}^{\perp}$) exchange interaction paths.}
\label{theory-fig} }
\end{figure}

\textcolor{black}{To corroborate the experimental findings with theoretical
calculations, we have performed spin-polarized GGA+U calculations
with a ferromagnetic (FM) arrangement of Cr spins. The local octahedral
environment splits the Cr-$d$ orbitals into low lying $t_{2g}$ triplets
and higher energy $e_{g}$ doublets maintaining a crystal field splitting
of $\sim$4.0~eV. The plot of spin-polarized density of states in
Fig.~\ref{theory-fig}~(a) reveals that the majority channel of
the Cr-$t_{2g}$ levels are completely filled with three valence electrons
and the minority channel of $t_{2g}$ and the $e_{g}$ orbitals are
completely empty. Within the FM calculation, the total moment per
formula unit, containing 6 Cr atoms, is calculated to be 18.0 $\mu_{B}$,
which further supports the 3+ charge state of Cr and is also consistent
with the experimental value of effective moment }($\mu_{{\rm eff}}$
= 3.84~$\mu_{B}$\textcolor{black}{). FM calculation gives the magnetic
moment per Cr site to be 2.90 $\mu_{B}$, as the rest of the moment
lies in the ligand sites (0.01 $\mu_{B}$/O). Inclusion of spin-orbit
coupling (SOC) in our calculation only lowers the energy of FM state
further by $74$~meV keeping the moment value per Cr similar to that
of without SOC calculations. The small value of orbital moments ($m_{l}$(Cr)$=0.04~\mu_{B}$)
in GGA+SOC+U calculations suggests SOC acts perturbative ($m_{l}/m_{s}\sim0.01$)
in the system and has negligible impact on stabilizing the ground
state. }
\textcolor{black}{{} }

\textcolor{black}{In order to understand the magnetic exchange interactions
among the large intrinsic Cr spins and for a quantitative estimation
of the Cr-Cr exchange couplings, we have calculated the symmetric
exchange interactions ($J$) by mapping the total energies of several
spin configurations to the Heisenberg spin model $\mathcal{H}_{{\rm sym}}=J_{ij}\bm{S_{i}}\cdot\bm{S_{j}}$
\cite{Ag3LiMn2O643}. Our calculations suggest that the nearest neighbor
(nn) interaction is ferromagnetic in nature with exchange strength
$J_{1}=0.22$~meV. The obtained further neighbor in-plane and out-of-plane
couplings (see Fig.~\ref{theory-fig}~(b) for connectivity) are
antiferromagnetic in nature with magnitudes $J_{1}^{\perp}=0.06$~meV,
$J_{2}=0.03$~meV, $J_{2}^{\perp}=0.05$~meV, $J_{3}=0.06$~meV
and $J_{3}^{\perp}=0.03$~meV, respectively. Due to large super-exchange
paths, the interaction strengths are relatively small which is also
presumed from the sufficiently low value of experimental Curie-Weiss
temperature ($\Theta_{{\rm CW}}^{{\rm Bulk}}$ }$\sim$\textcolor{black}{{}
4.54(2)~K or} $\Theta_{{\rm CW}}^{{\rm NMR}}$\textcolor{black}{{}
}$\sim$\textcolor{black}{{} 2.88(11)~K). Clearly, the FM exchange
interaction $J_{1}$, corresponding to the nn super-exchange coupling,
dominates over the further neighbor anti-ferromagnetic couplings.
To explore the possibility of spin canting in the system, we have
also considered the anti-symmetric part of the spin Hamiltonian $\mathcal{H}_{{\rm asy}}=\sum_{ij}\bm{D}_{ij}\cdot(\bm{S}_{i}\times\bm{S}_{j})$,
where $\bm{D}$ is the Dzyaloshinskii-Moriya (DM) interaction parameter
\cite{dm_ref}. The calculated values of nn DM interaction is of the
order of $\sim10^{-3}$~meV. A small value of SOC strength results
in the suppressed value of DM parameter. Taken together the nn FM
interaction with a number of AFM interactions and small anti-symmetric
DM interaction, the calculated value of }$\Theta_{{\rm CW}}$\textcolor{black}{{}
\textasciitilde{} 0.3~K, which is reasonably small and further affirms
a ferromagnetic ground state.}\textbf{\textcolor{black}{{} }}\textcolor{black}{However,
it must be noted that the calculated value of }$\Theta_{{\rm CW}}$\textcolor{black}{{}
depends on the chosen exchange configuration and could vary between
\textasciitilde{} 0.3~(}$\Theta_{{\rm CW}}^{{\rm min}}$\textcolor{black}{:
summing up all the $J$'s)~ \& 12~K~(}$\Theta_{{\rm CW}}^{{\rm max}}$\textcolor{black}{:
only nearest neighbor). Apparently, the experimentally estimated value
of} $\Theta_{{\rm CW}}^{{\rm NMR}}$\textcolor{black}{{} }$\sim$\textcolor{black}{{}
2.88(11)~K is less than the computed value of }$\Theta_{{\rm CW}}^{{\rm max}}$\textcolor{black}{{}
}$\sim$12~K, implying the relevance of additional antiferromagnetic
couplings beyond the nearest neighbor ferromagnetic coupling (\textcolor{black}{$J_{1}=0.22$~meV}).\textcolor{black}{{}
Meisso $et$ $al.$ \cite{Kagome order 55} sketched a phase diagram
based on the lattice symmetries of a perfect kagomé lattice for the
identification of a most probable ground state in a large parameter
space of Heisenberg model with the intra-planar kagomé couplings $J_{1}$-$J_{2}$-$J_{3}$-$J_{3}^{,}$,
which suggests the stabilization of a rather complex ground state
for LCPO and places it in the proximity of a ferromagnetic and spiral
state. Nevertheless, in view of the skewed $C$$_{6}$ symmetry with
additional interlayer couplings (as is the case with LCPO) the fate
of the ground state selection is beyond the purview of the present
model and calls for further investigations.}

\subsection{NMR Spectroscopy}

\begin{figure}
\begin{centering}
\includegraphics[scale=0.335]{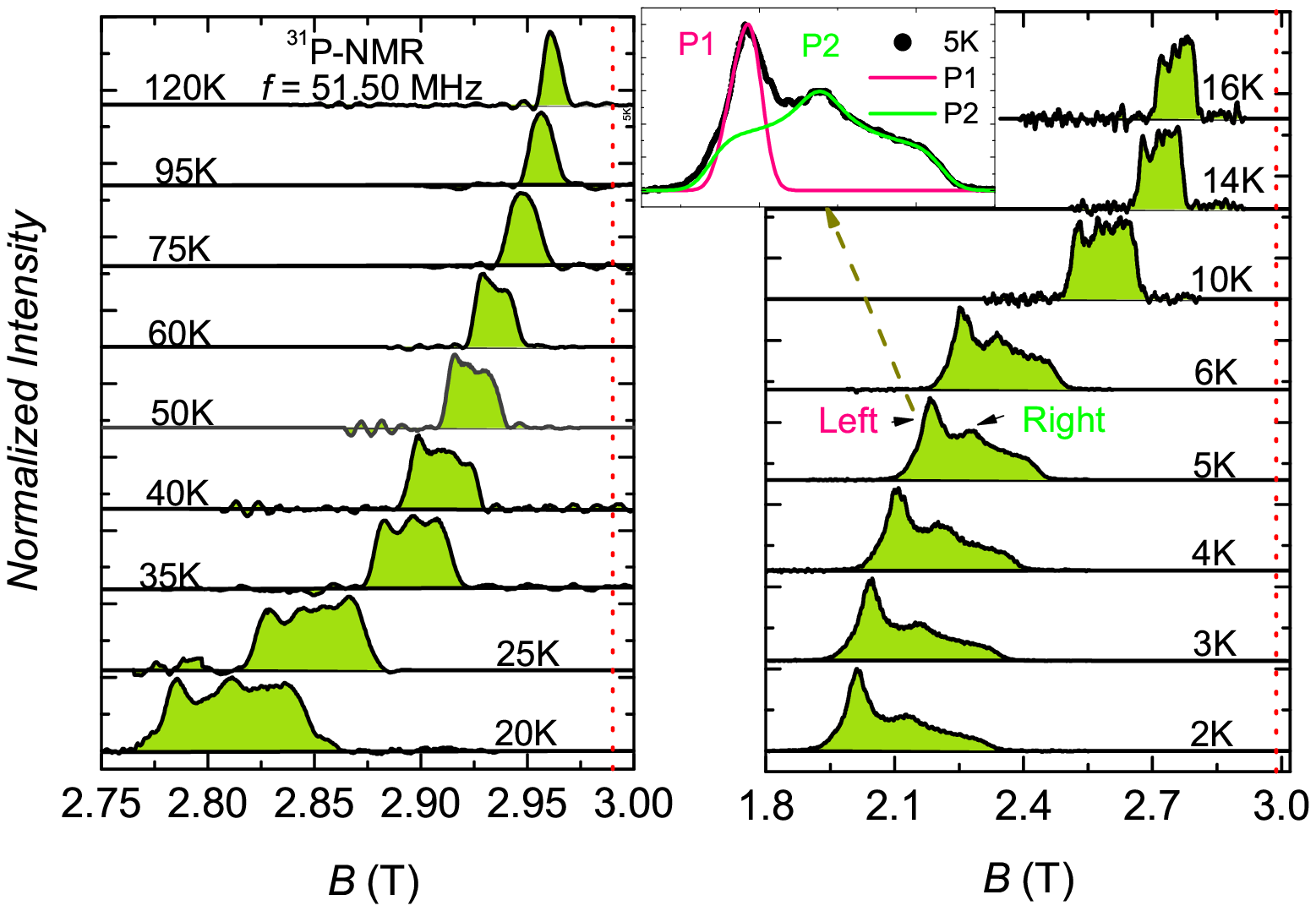} 
\par\end{centering}
\caption{\label{fig:NMR Spectra} $^{31}$P-NMR spectra measured at NMR frequency
$f$ = 51.50~MHz. Vertical dashed line depicts the $^{31}$P-NMR
reference field. The peaks marked with ``Left'' and ``Right''
are originating from two different phosphorous sites P(1) and P(2),
see text. Inset: $^{31}$P-NMR spectrum at 5~K fitted with two NMR
lines; P1~(Left peak) and P2~(Right peak).}
\end{figure}

The LCPO offers two potential NMR sensitive nuclei, $^{7}$Li ($I$
= 3/2 \& $\frac{\gamma}{2\pi}$ = 16.546~MHz/T) and $^{31}$P ($I$
= 1/2 \& $\frac{\gamma}{2\pi}$ = 17.235~MHz/T) to probe the magnetism
ensuing from the kagomé network formed by the Cr$^{3+}$ ions. $^{7}$Li
was found to be poorly coupled to the kagomé network because of a
weak hyperfine coupling, resulting from the relatively extended pathways.
We then switched to $^{31}$P - NMR to probe the static susceptibility
and relaxation rate measurements. LCPO being a low-$J$~( < 1~K)
material with a small saturation field \textasciitilde{} 1~T, it
was challenging \cite{NMR Area44} to address the ground state properties
of the material by employing NMR as a probe for magnetic fields comparable
to the energy scale set by the exchange-coupling. In lieu of that
we investigated the effect of magnetic field on the ground state properties
for $B$ $\geq$ 3~T. The $^{31}$P - NMR spectra were collected
at fixed frequencies ($f$ = 51.50~MHz, 120.51~MHz, and 215.50~MHz)
by sweeping the magnetic field using the Hahn-echo. Figure \ref{fig:NMR Spectra}
shows the $^{31}$P - NMR spectra measured in the temperature range
120 - 2~K for the NMR frequency $f$ = 51.50~MHz. Structurally there
are two distinct P-sites, P(1) \& P(2), in a unit cell of LCPO with
the Wyckoff positions 4g~(2/3, 1/3, z) and 12d~( x, y, z) respectively,
but in the $T$-range 120 - 75~K, we observed only one $^{31}$P
- NMR line with a nearly Gaussian line-shape because the other $^{31}$P
- NMR line was also buried under the same line as a result of having
a nearly similar NMR shift \cite{NMR Area44}. Upon further lowering
the temperature, the $^{31}$P - NMR line shape starts developing
structure and progressively becomes highly asymmetric; nevertheless
below about 10~K the spectral shape can largely be interpreted as
a superposition of two NMR lines: \textit{``Left}'' - line \& \textit{``Right}''
- line, see Fig. \ref{fig:NMR Spectra}. As evident from Fig. \ref{Structure},
phosphorus atoms located at the sites: 4g, $i.e.,$ P(1), and 12d,
$i.e.,$ P(2), are coupled to three and two Cr - atoms, respectively,
therefore $^{31}$P - NMR signal corresponding to P(1) is expected
to be shifted more in comparison to P(2) with respect to the NMR reference-line,
and also the $^{31}$P -NMR spectrum arising from P(1) is anticipated
to be less anisotropic in comparison to P(2) as a result of being
situated at a more symmetric position in the unit-cell. Owing to their
atomic weight percentage (4 : 12) in a unit cell of LCPO, both P(1)
and P(2) are expected to maintain an intensity ratio of 1 : 3 in $^{31}$P
- NMR spectrum. With this notion, we chose a spectrum collected at
5~K (inset of Fig.  \ref{fig:NMR Spectra}) and the $^{31}$P - NMR
shift was fitted using Eq. (1) and the obtained results are summarized
in Table \ref{Spectrum FIT}.

$K$ = $K_{iso}$ + $K_{ax}$(3cos$^{2}\theta$ - 1) + $K_{ani}$~sin$^{2}\theta$
cos$2\phi$, ~~~~~~(1)

where $K_{iso}$, $K_{ax}$, and $K_{ani}$ are the isotropic, axial,
and anisotropic components of the Knight shift, respectively. While
$\theta$ and $\phi$ are spherical polar angles defining the orientation
of $B$ with respect to the shift tensor axes. 

\begin{table}
\caption{\label{Spectrum FIT} Fitting parameters for $^{31}$P-NMR spectrum
collected at 5~K.}

\begin{tabular}{|>{\centering}p{0.75cm}|>{\centering}p{1.2cm}|>{\centering}p{1.2cm}|>{\centering}p{1.2cm}|>{\centering}p{1.2cm}|>{\centering}p{1.5cm}|c|}
\hline 
Line  & $K_{iso}$(\%)  & $K_{ani}$(\%)  & $K_{ax}$(\%)  & Width (kHz)  & Area (arb. units) & Ratio\tabularnewline
\hline 
\hline 
P1  & 36.86 & \ensuremath{\simeq}~0 & 0.93 & 300.05 & 0.038 & 1\tabularnewline
\hline 
P2  & 30.97 & 4.3 & 4.2 & 661.05 & 0.116 & 3.05\tabularnewline
\hline 
\end{tabular}
\end{table}

As can be seen from the inset of Fig. \ref{fig:NMR Spectra} and the
tabulated results, the\textit{ ``Left}'' - peak (P1) is nearly a
Gaussian, less anisotropic ($K_{ani}$ \ensuremath{\simeq}~0~\%)
in nature (as P1 is located at the 3-fold axis of rotational symmetry),
while the \textit{``Right}'' - peak (P2) of the spectrum is extremely
broadened, highly anisotropic ($K_{ani}$ = 4.3~\%) and the ratio
of calculated integrated intensity for P1 and P2 is in excellent agreement
with the expected ratio \textasciitilde{} 1 : 3.

Based on the fitting results, in line with the underlying presumptions,
we can unambiguously assign the \textit{``Left}'' - peak of $^{31}$P
- NMR spectrum to be originating from the phosphorus - P(1) (shown
by line P1) and the \textit{``Right}'' - peak from the phosphorus
- P(2) (shown by line P2).

\begin{figure}
\begin{centering}
\includegraphics[scale=0.35]{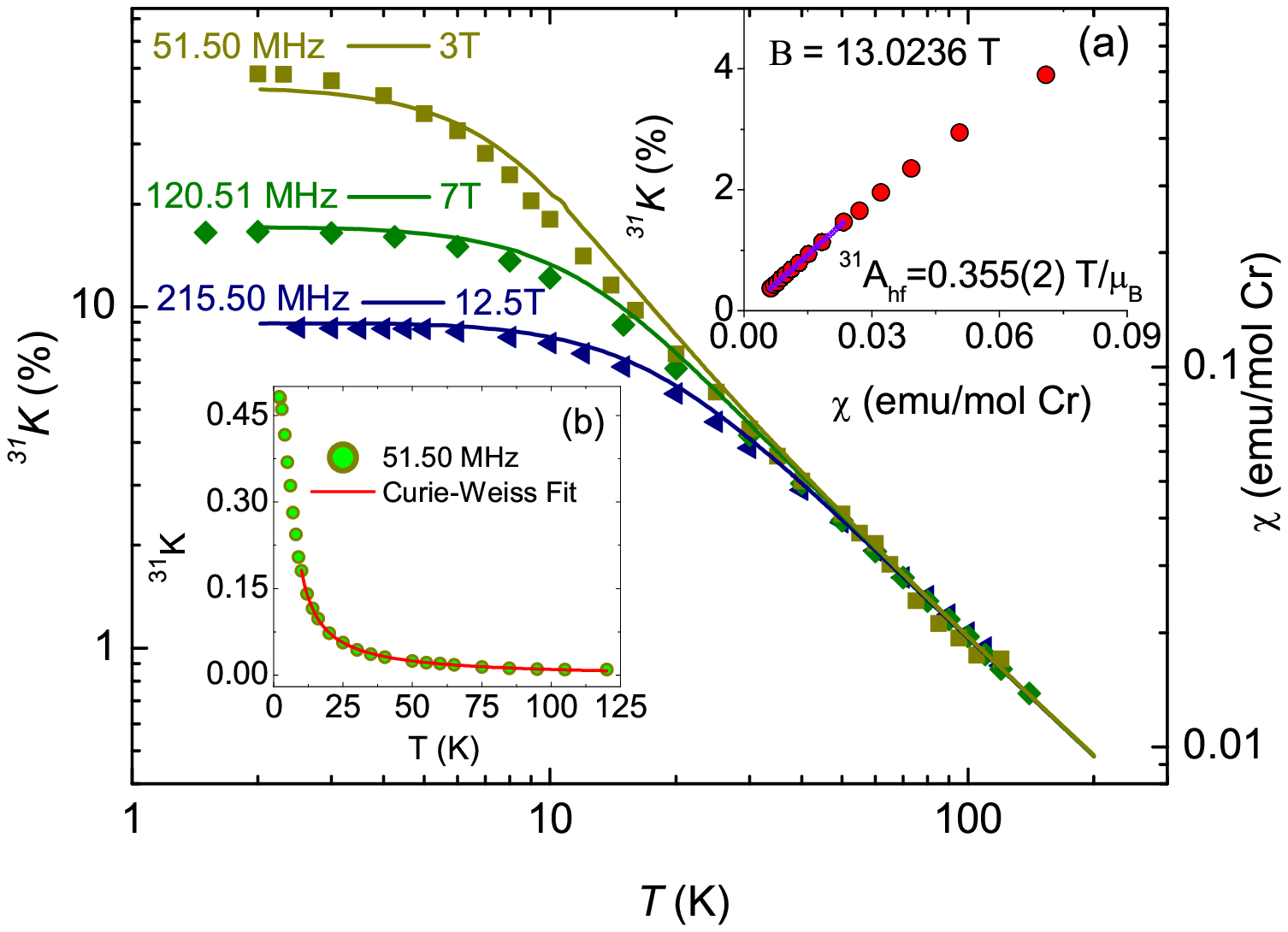} 
\par\end{centering}
\caption{\label{fig:NMR-shift} Temperature dependence of $^{31}$P - NMR line
shift (symbols) measured at three different NMR frequencies: $f$
= 51.50~MHz (3~T), 120.51~MHz~(7~T), and 215.50~MHz~(12.5~T)
(left y - axis) for line P1. Variation of bulk susceptibility (lines)
with temperature measured at 3~T, 7~T and 12.5~T (right y - axis).
Insets: (a) $K$ - $\chi$ plot for line P1, and (b) $^{31}$P - NMR
line shift (filled circle) measured at $f$ = 51.50~MHz and the line
(red) is a fit to the Curie-Weiss type equation, described in text.}
\end{figure}

It was not possible to isolate the contributions of two peaks from
the observed NMR line and thus the deduction of isotropic ($K_{iso}$)
and anisotropic ($K_{ani}$) NMR shifts in the paramagnetic region,
so we plotted the Knight shift ($K$) for peak P1 by taking the peak
position. Figure \ref{fig:NMR-shift} shows the $^{31}$P-NMR shift
($K$), for P1, as a function of temperature in the $T$-range 1.5
- 120~K at three different NMR frequencies. The NMR shift measurements
provide a way to extract the intrinsic susceptibility (free from extraneous
contributions) through the relation: $^{31}$$K$ = \textcolor{black}{$K_{0}$+
}$(A_{{\rm hf}}$$/N_{A}\mu_{B})$$\chi_{{\rm local}}$\textcolor{black}{,
where $K_{0}$, }$A_{{\rm hf}}$\textcolor{black}{{} and }$\chi_{{\rm local}}$
are the temperature independent chemical shift, hyperfine - coupling
constant, and the local spin susceptibility, respectively. As evident
from Fig. \ref{fig:NMR-shift}, the NMR shift ($^{31}$$K$) nicely
follows the bulk susceptibility in the entire temperature range, therefore
confirming the intrinsic nature of powder susceptibility data and
its variation with external magnetic field. The suppression of bulk
susceptibility/NMR shift as a function of magnetic field is in accordance
with the magnetic isotherm data. The hyperfine coupling constant ($A_{{\rm hf}}$)
was determined for P(1)\textbf{ }by measuring the shift at the center
position with a fixed field ($B$ = 13.0236~T) and estimated to be
0.355(2)~T/$\mu_{B}$, shown in the inset (a) of Fig. \ref{fig:NMR-shift}.
\textcolor{black}{Fitting the shift ($^{31}K$) obtained at 51.50~MHz
to the Curie-Weiss law {[}$^{31}K$ = $K$$_{0}$+ $C$/(T-$\varTheta$){]}
yields a more refined value of $\varTheta$ = 2.88(11)~K or $J$
= 3}$\varTheta$/(4$S$($S$+1) \textasciitilde{} 0.6~K\textcolor{black}{{}
(compared to bulk susceptibility) with $K_{0}$ = -0.0037(4), and
$C$ = 1.32(2)~K, see inset (b) of }Fig. \ref{fig:NMR-shift}\textcolor{black}{.
The offset (\textasciitilde 1.5~K) in the value of $\varTheta$
determined from the bulk susceptibility and NMR results could be the
upshot of some extrinsic contributions affecting the paramagnetic
behavior of bulk magnetization. As the ordering temperature, $T_{c}$
= 2.7~K (determined from the $C_{p}$ data), and the mean-field temperature,
$\varTheta$ = 2.88(11)~K, (from the $^{31}$P-NMR line shift), }are
nearly the same, implying that frustration plays no role in selecting
the ground state.

\begin{figure}[b]
\centering{}\includegraphics[scale=0.35]{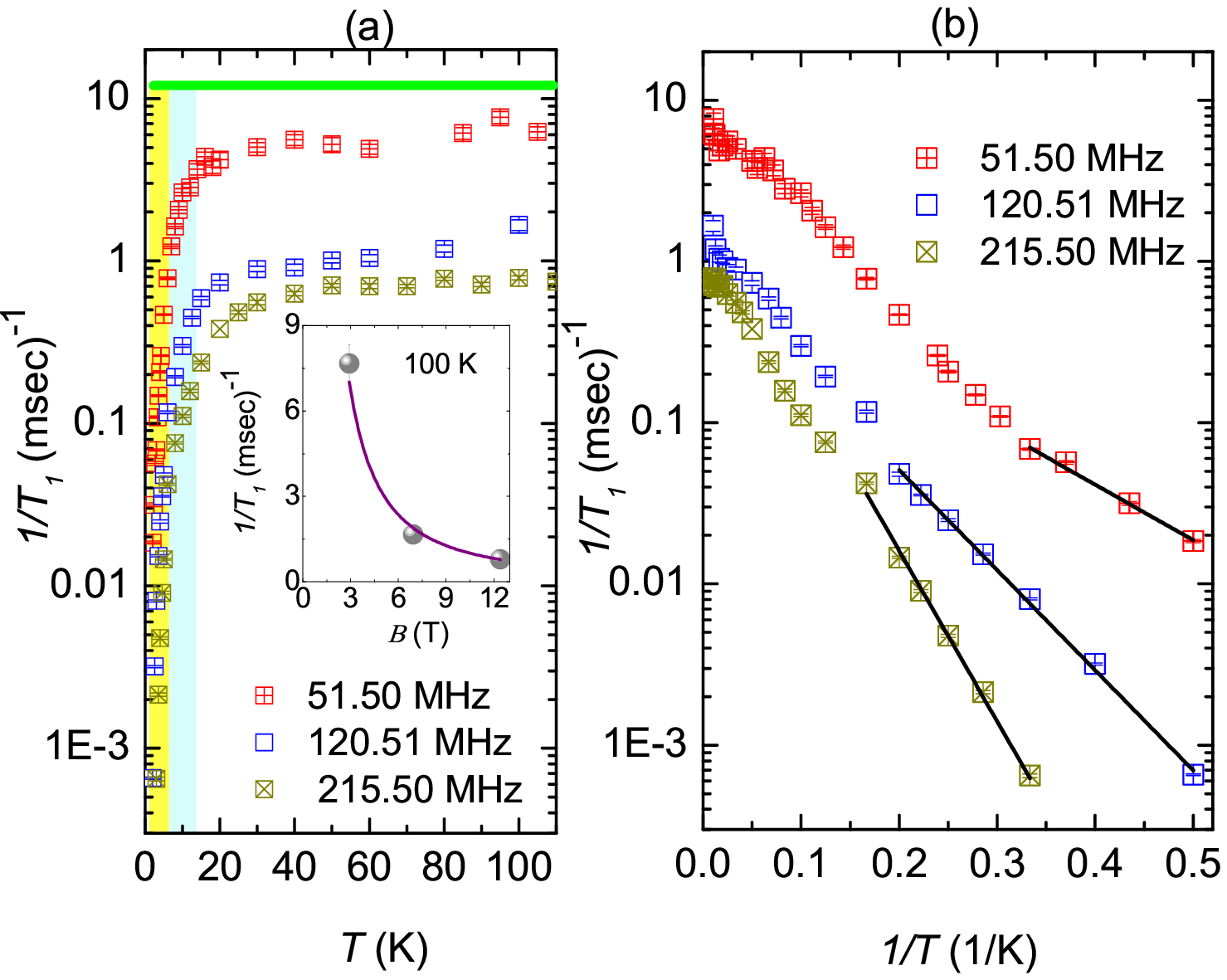}\caption{\label{fig:Spin-lattice-relaxation-rate} Spin-lattice relaxation
rate ($1/T_{1}$) measured at the peak position of the P1 - NMR line
at three different fields. (a) Semi - log plot of $1/T_{1}$ versus
$T$ with the Moriya - limit shown by the horizontal green line. Inset
shows the field variation of $1/T_{1}$ (filled circle) at $T$ =
100~K and the line (purple) is a fit to the power - law: $1/T_{1}$
$\propto$ $B$$^{-n}$. (b) $1/T_{1}$ versus $1/T$ in log-log scale;
black solid lines are the fits to a function (see text).}
\end{figure}

Now we focus our attention to understand the effect of magnetic field
on the spin dynamics and for that purpose the spin-lattice relaxation
rate ($1/T_{1}$) measurements were performed by employing the saturation
recovery method at the peak position of line P1, \textit{``Left}''-peak
of $^{31}$P-NMR spectrum, in the $T$- range 1.5 - 130~K and at
three NMR frequencies: $f$ = 51.50~MHz~(3~T), 120.51~MHz~(7~T),
and 215.50~MHz~(12.5~T). The saturation recovery data were fitted
to the equation: $M_{z}$(t) = $M_{0}$(1 - $A$$e^{-t/T_{1}}$) and
the extracted values of spin-lattice relaxation rates ($1/T_{1}$)
are plotted against temperature ($T$), see Fig. \ref{fig:Spin-lattice-relaxation-rate}~(a).

$1/T_{1}$ shows a marked frequency dependence even in the paramagnetic
regime ($T$ $\gg$$\Theta_{{\rm CW}}$). For the lowest frequency
$f$ = 51.50~MHz, it attains a maximum value \textasciitilde{} 8~(msec)$^{-1}$
at 100~K and for the higher frequencies it reduces in a non-linear
fashion in the paramagnetic region, see Fig. \ref{fig:Spin-lattice-relaxation-rate}~(a).
In the high temperature region, where the $^{31}$P - NMR nuclear
relaxation is supposed to be dominated by the fluctuations of the
localized moments of the chromium ions, $1/T_{1}$ can be estimated
using the Moriya relation \cite{Moriya45}: $1/T_{1}|_{{\rm Moriya}}$
= $(2\pi)^{1/2}$($A_{{\rm hf}}$$/\hbar)^{2}$$Z_{1}S(S+1)/3\omega_{ex};$
and $\omega_{ex}$ = ($6k_{B}^{2}\varTheta^{2}/Z_{2}S(S+1)\hbar^{2}$)$^{1/2}$,

where\textcolor{black}{{} }$A_{{\rm hf}}$, $Z_{1}$, $\omega_{ex}$,
$\Theta_{{\rm CW}}$, and $Z_{2}$ are hyperfine coupling constant,
number of magnetic ions connected to nuclei being probed, exchange
frequency, Curie-Weiss temperature, and immediate neighbors of a magnetic
ion, respectively.

\begin{table}[H]
\caption{\label{Moriya} Parameters used in estimating the spin-lattice relaxation
rate $1/T_{1}|_{{\rm Moriya}}$ using the Moriya relation.}

\begin{tabular}{|c|c|c|c|c|c|}
\hline 
$Z_{1}$  & $Z_{2}$  & $\Theta_{{\rm CW}}$~($K$)  & $g$  & $1/T_{1}|_{{\rm Moriya}}$ (msec)$^{-1}$  & Peak\tabularnewline
\hline 
\hline 
3  & 4  & 3  & 1.94  & 11.60 & \textcolor{magenta}{P1}\tabularnewline
\hline 
2  & 4  & 3  & 1.94  & \textasciitilde 13.0 & \textcolor{lime}{P2}\tabularnewline
\hline 
\end{tabular}
\end{table}

 The calculated Moriya limit $1/T_{1}|_{{\rm Moriya}}$ for both the
peaks P1 \& P2 is tabulated in Table \ref{Moriya} and the relaxation-rate
obtained for P1 (at $f$ = 51.50~MHz \& $T$ = 100~K) is seemingly
very close to the Moriya-limit, see Fig. \ref{fig:Spin-lattice-relaxation-rate}~(a).
On the other hand, a field induced reduction of $1/T_{1}$ seen above
20~K ($\gg$$\Theta_{{\rm CW}}$) is most likely occurring due to
a relaxation mechanism which involves the participation of lattice
vibrational modes; a similar trend was also noticed in the $1/T_{1}$
behavior of sister material LFPO \cite{Edwin LFPO15}. Inset of Fig.
\ref{fig:Spin-lattice-relaxation-rate}~(a) depicts the variation
of $1/T_{1}$ with applied field at $T$ = 100~K and a fit to the
expression: $1/T_{1}$ = $A$$B$$^{-n}$ outputs in a good fit with
$n$ = 1.5~(1). Such kind of relaxation behavior \cite{Orbach 57}
is governed by the direct process, $n$ = 2 (non - Kramers ion) and
4 (Kramers ion), and has been discussed in the context of rare-earth
salts at low temperatures with a possible extension to 3$d$ material,
but its relevance in the high temperature limit is questionable. 

\begin{figure}
\begin{centering}
\includegraphics[scale=0.35]{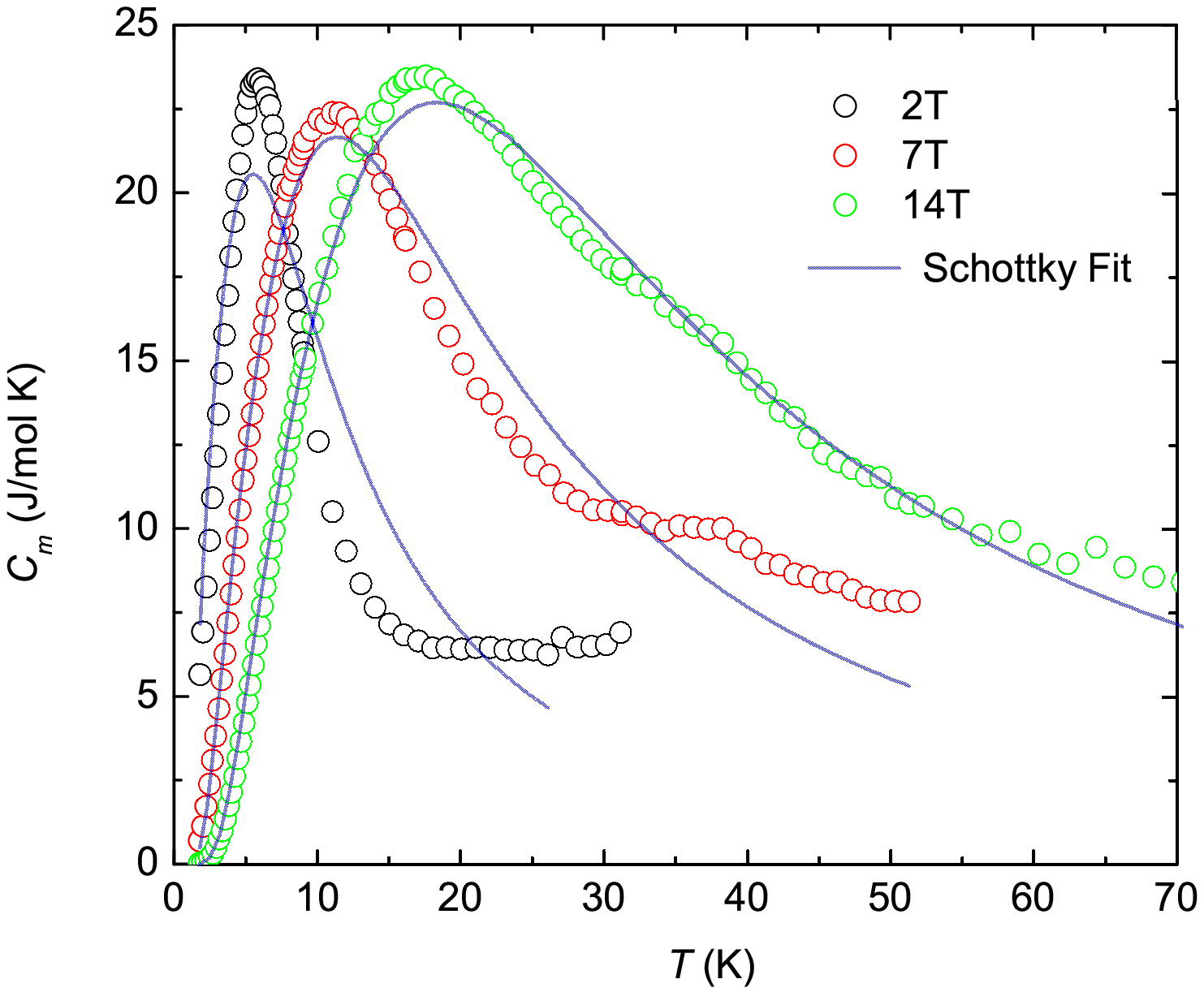}
\par\end{centering}
\caption{\label{fig:Schottky-fitting} Temperature dependence of magnetic specific
heat (open circle) and its fitting (solid line) to the four-level
Schottky system at the selected magnetic fields.}
\end{figure}

Now, we focus to understand the low - $T$ behavior of $1/T_{1}$.
$1/T_{1}$ in polarized or partially polarized state of a magnet is
expected to have a gapped behavior. $1/T_{1}$ data fitted to an equation
of the Arrhenius form: $1/T_{1}$ $=A$exp$(-\Delta_{N}/k_{B}T)$
yields a gap, $\Delta_{N}$ ($T$ - range: \textasciitilde{} 1.5 -
3~K for 3~T and \textasciitilde{} 1.5 - 6.5~K for 7 \& 12.5~T),
see Fig. \ref{fig:Spin-lattice-relaxation-rate}~(b). 
\begin{figure}[t]
\begin{centering}
\includegraphics[scale=0.35]{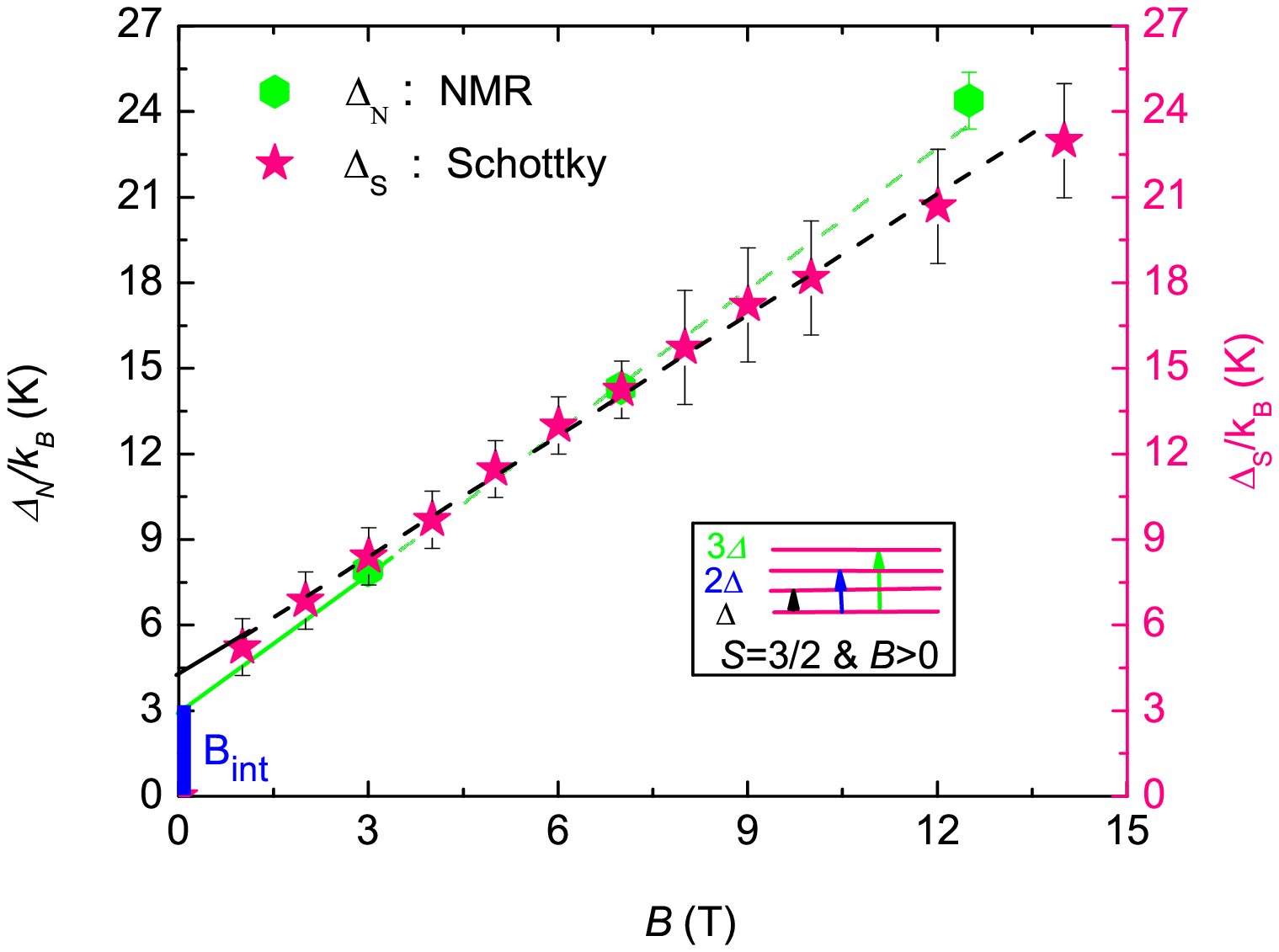}
\par\end{centering}
\caption{\label{fig:NMR=000026Specific heat gaps} Field dependence of gaps
estimated using NMR (left $y$ - axis: ($\Delta_{N}/k_{B}$) and specific
heat (right $y$ - axis: $\Delta_{S}/k_{B}$) data. A schematic of
the Zeeman splitting scheme for $S$ = 3/2 in paramagnetic region
is also shown. The green (NMR) and black (Schottky) dashed lines are
fits to a straight line and the vertical thick line (blue) depicts
the size of internal field. }
\end{figure}

\textcolor{black}{In order to gain more insight about the nature and
origin of this gap, we further analyzed the specific heat data}\textcolor{green}{{}
}by modeling the Schottky anomaly (Fig. \ref{fig:Schottky-fitting})
and compared the results with gap obtained using NMR data, shown in
Fig. \ref{fig:NMR=000026Specific heat gaps}. 

In paramagnetic region and under non-zero external magnetic field
($B$ $\neq0$), a $S$ = 3/2 system would have equally spaced (2$S$+1)
= 4 energy levels with the lowest and highest energies being 0 and
3$\Delta$, respectively, shown in Fig. \ref{fig:NMR=000026Specific heat gaps}.
The magnetic specific heat, $C_{m}$, was fitted to a four level system
with the following expression \cite{Heat capacity46}:

\[
C_{m}=\frac{R}{k_{B}^{2}T^{2}}\left[\frac{\sum_{i=0}^{3}\epsilon_{i}^{2}e^{-(\epsilon_{i}/k_{B}T)}}{\sum_{i=0}^{3}e^{-(\epsilon_{i}/k_{B}T)}}-\left\{ \frac{\sum_{i=0}^{3}\epsilon_{i}e^{-(\epsilon_{i}/k_{B}T)}}{\sum_{i=0}^{3}e^{-(\epsilon_{i}/k_{B}T)}}\right\} ^{2}\right],
\]

where $\epsilon_{i}$ denotes the energy of $i^{th}$ spin-level and
the gap is defined as $\Delta_{i}=\epsilon_{i}-\epsilon_{0}$ and
the other notations have their usual meanings.

The Schottky fitting deviates from the experimental data below 7~T
and results in a rather poor fit at 2~T, see Fig. \ref{fig:Schottky-fitting}.
This result is suggesting that a gradual reduction observed in the
Schottky peak as a function of field, Fig. \ref{Heat capacity}~(b),
is a consequence of magnetic correlations, which persist above the
mean-field temperature. The Schottky gap, $\Delta_{S}$, estimated
by modeling the magnetic specific heat data not only nicely coincides
with the NMR gap~($\Delta_{N}$) in a wider field range (deviates
slightly at the higher fields \cite{Schottky 60}), but also linearly
scales with the external magnetic field, and in $B$$\rightarrow0$
limit both the gaps ($\Delta_{S}$ \& $\Delta_{N}$) attain a non-zero
value of intercept $\sim$ 4.1~(1)~K \& 2.8~(5)~K, respectively,
see Fig. \ref{fig:NMR=000026Specific heat gaps}. \textcolor{black}{Apparently,
the gap size estimated using these two techniques: bulk~(Specific
heat) and local ($^{31}$P - NMR) probes is quite consistent}\textit{\textcolor{black}{{}
}}\textcolor{black}{and we believe that the gap size estimated using
NMR is much more precise as the NMR transitions are constrained by
the selection rules (}$\Delta m$\textcolor{black}{{} = 1) imposed
by the magnetic dipole transitions, on the contrary, the} specific
heat yields an average value of gap,\textcolor{black}{{} therefore
it sets an upper-bound to the gap size.}\textit{\textcolor{black}{{}
}}Interestingly, the gap, $\Delta_{N}$, estimated by NMR technique
in $B$$\rightarrow0$ limit has an intercept value close to the mean-field
temperature scale \textasciitilde{} 2.88~K, which essentially implies
the size of the internal field developed as a result of magnetic ordering.
For fields above $B$ $\geq$ 3~T, the external field takes over
and the gap scales linearly with the applied field. 

\section{Conclusion}

In summary, we have investigated the structural and physical properties
of a kagomé material, LCPO, using x-ray diffraction \& neutron diffraction,
thermodynamic, and $^{31}$P-NMR measurements. Even though LCPO is
isostructural to LFPO, the ground state of LCPO is in stark contrast
to LFPO and does not show the effect of frustration (offered by the
kagomé network) and relatively stronger quantum fluctuations than
LFPO as a result of reduced spin quantum number $S=3/2$. The LCPO
stabilizes a ferromagnetic like ground state with $T$$_{c}$ = 2.7~K,
and the short-range correlations are found to survive above the mean-field
temperature $\Theta_{{\rm CW}}$ \textasciitilde{} 3~K. The presence
of additional couplings ($2^{nd}$ and $3^{rd}$) turn the material
away from a pure nearest Heisenberg kagomé ferromagnet. This structurally
clean kagomé material is quite a rare example of an insulating ferromagnet,
where the ferromagnetism is a spin-off of nn exchange interaction
rather than the DM and dipolar interactions.\textbf{ }The presence
of highly degenerate flat bands in kagome lattice makes it conducive
to hold the ferromagnetic state and the absence of DM interaction
in LCPO could again protect the degeneracy of flat bands and thus
the zero-energy excitations. In order to understand the role of these
bands on the elementary excitatons and in turn on the ground state,
inelastic neutron scattering experiments could be of paramount importance. 
\begin{acknowledgments}
RK thanks TIFR Mumbai for facilitating with the experimental facilities
for conducting the initial research work. \textcolor{black}{A.C acknowledges
Indian Institute of Technology, Kanpur and Science and Engineering
Research Board (SERB) National Postdoctoral Fellowship (PDF/2021/000346),
India for financial support.} P L thanks Sudhindra Rayaprol for conducting
initial neutron diffraction measurements. We thank CC-IITK for providing
the high performance computing facility. This work is partly supported
by the JSPS Grant-in Aid for Scientific Research (Grant Nos 19H01832,
20K20892, 21H01035, 22H04458) and the ISSP Institutional Collaborative
Research Program. 
\end{acknowledgments}


\begin{thebibliography}{10}
\bibitem[1]{Balenta1} L. Balents, Nature (London) \textbf{464}, 199
(2010).

\bibitem[2]{Emergent excitations 2} B. Bauer, L. Cincio, B. P. Keller,
M. Dolfi, G. Vidal, S. Trebst, and A. Ludwig, Nat. Commun. \textbf{5},
5137 (2014).

\bibitem[3]{Heydite-3} D. Boldrin, B. Fåk, M. Enderle, S. Bieri,
J. Ollivier, S. Rols, P. Manuel, and A. S. Wills, Phys. Rev. B\textbf{
91}, 220408(\textbf{R}) (2015).

\bibitem[4]{Cu(bdc)-4} R. Chisnell, J. S. Helton, D. E. Freedman,
D. K. Singh, R. I. Bewley, D. G. Nocera, and Y. S. Lee, Phys. Rev.
Lett. \textbf{115}, 147201 (2015)

\bibitem[5]{Cu(bdc)-5 Cp} M. Hirschberger, R. Chisnell, Y. S. Lee,
and N. P. Ong, Phys. Rev. Lett. \textbf{115}, 106603 (2015).

\bibitem[6]{Helton Kagome6} J. S. Helton, K. Matan, M. P. Shores,
E. A. Nytko, B. M. Bartlett, Y. Yoshida, Y. Takano, A. Suslov, Y.
Qiu, J.-H. Chung, D. G. Nocera, and Y. S. Lee, Phys. Rev. Lett. \textbf{98},
107204 (2007).

\bibitem[7]{Broholom Fractaionalized excitations7} T.-H. Han, J.
S. Helton, S.Chu, D. G. Nocera, J. A. Rodriguez- Rivera, C. Broholm,
and Y. S. Lee, Nature (London)\textbf{ 492}, 406 (2012).

\bibitem[8]{Norman RMP8} M.\LyXThinSpace R. Norman, Rev. Mod. Phys.
\textbf{88}, 041002 (2016).

\bibitem[9]{Hida S1-9} K. Hida, J. Phys. Soc. Jpn. \textbf{69}, 4003
(2000).

\bibitem[10]{Coupled cluster HAK for S10} O. Götze, D. J. J. Farnell,
R. F. Bishop, P. H. Y. Li, and J. Richter, Phys. Rev. B\textbf{ 84},
224428 (2011).

\bibitem[11]{HSS ANsatz 11} Wei Li, Andreas Weichselbaum, Jan von
Delft, and Hong-Hao Tu, Phys. Rev. B \textbf{91}, 224414 (2015).

\bibitem[12]{Spin-S Magnetization12} H. Nakano, and T. Sakai, J.
Phys. Soc. Jpn. \textbf{84}, 063705 (2015).

\bibitem[13]{Spin ordered gs for Cr13} Tao Liu, Wei Li, and Gang
Su, Phys. Rev. E \textbf{94}, 032114 (2016).

\bibitem[14]{Changlani NaTiO14} A. Paul, Chia-Min Chung, T. Birol,
and H. J. Changlani, Phys. Rev. Lett. \textbf{124}, 167203 (2020).

\bibitem[15]{Edwin LFPO15} E. Kermarrec, R. Kumar, G. Bernard, R.
Hénaff, P. Mendels, F. Bert, P. L. Paulose, B. K. Hazra, B. Koteswararao,
Phys. Rev. Lett. \textbf{127}, 157202 (2021).

\bibitem[16]{Hifa HSS16} K. Hida, J. Phys. Soc. Jpn. \textbf{69},
4003 (2000).

\bibitem[17]{Hao arxiv QQSN 17} H. Yao, L. Fu, and X.-L. Qi, arXiv:1012.4470
(2010).

\bibitem[18]{TOp=00003D000026 criticality AKLT18} W. Li, S. Yang,
M. Cheng, Z.-X. Liu, and H.-H. Tu, Phys. Rev. B \textbf{89}, 174411
(2014).

\bibitem[19]{NematicS12@19} T. Picot and D. Poilblanc, Phys. Rev.
B \textbf{91}, 064415 (2015).

\bibitem[20]{Quasi2dSCGO-20} A. P. Ramirez, G. P. Espinosa, A. S.
Cooper, Phys. Rev. Lett. \textbf{64}, 2070 (1990).

\bibitem[21]{SGCOdilutionGlass21} B. Martí­nez, F. Sandiumenge, A.
Rouco, A. Labarta, J. Rodrí­guez-Carvajal, M. Tovar, M.T. Causa, S.
Galí­, and X. Obradors, Phys. Rev. B \textbf{46}, 10786 (1992).

\bibitem[22]{Kerenmusr-22} A. Keren, Y. J. Uemura, G. Luke, P. Mendels,
M. Mekata, and T. Asano, Phys. Rev. Lett. \textbf{84}, 3450 (2000).

\bibitem[23]{Cp SCGO23} A. P. Ramirez, B. Hessen, and M. Winklemann,
Phys. Rev. Lett. \textbf{84}, 2957 (2000).

\bibitem[24]{SCGOSL 24} D. Bono, L. Limot, and P. Mendels, Low Temp
Phys. \textbf{31}, 704 (2005).

\bibitem[25]{LeeGlass SCGO25}S.-H. Lee, C. Broholm, G. Aeppli, A.
P. Ramirez, T. G. Perring, C. J. Carlile, M. Adams, T. J. L. Jones
and B. Hessen, Europhys. Lett. \textbf{35},127 (1996).

\bibitem[26]{GlassSCGO26} I. Klich, S.-H. Lee, K. Iida, Nat Commun.
\textbf{5}, 3497 (2014).

\bibitem[27]{SCGO topoglass27} J. Yanga , A. Samarakoon, S. Dissanayake,
H. Ueda, I. Klicha , K. Iida, D. Pajerowski , N. P. Butch, Q. Huang,
J. R. D. Copley, and S.-H. Lee, Proc. Natl. Acad. Sci. \textbf{112},
11519 (2015).

\bibitem[28]{Poisson CrFe 28} S. Poisson, F. d'Yoire, NGuyen-Huy-Dung,
E. Bretey, and P. Berthet, J. Solid State Chem. \textbf{138}, 32 (1998).

\bibitem[29]{LVPOprep29} Q. Kuang, J. Xu, Y. Zhao, C. Chen, and L.
Chen, Electrochim. Acta \textbf{56}, 2201 (2011).

\bibitem[30]{LVPOprep30} M. Onoda and M. Inagaki, J. Phys. Soc. Japan
\textbf{80}, 084801 (2011).

\bibitem[31]{LVPONMR31} M. Onoda and S. Ikeda, J. Phys. Soc. Jpn.
\textbf{82}, 074801 (2013).

\bibitem[32]{LVPONMR32} M. Onoda and S. Ikeda, J. Phys. Soc. Jpn.
\textbf{88}, 034709 (2019).

\bibitem[33]{LVPONMRTheory33} M. Onoda and S. Takada, J. Phys. Soc.
Jpn. \textbf{89}, 034002 (2020).

\bibitem[34]{V jaro-34} D. Grohol, Q. Huang, B. H. Toby, J. W. Lynn,
Y. S. Lee, and D. G. Nocera, Phys. Rev. B \textbf{68}, 094404 (2003).

\bibitem[35]{Fe Jaro wills35} A. S. Wills, A. Harrison, C. Ritter,
and R. I. Smith, Phys. Rev. B \textbf{61}, 6156 (2000).

\bibitem[36]{musrKerenjaro36} A. Keren, K. Kojima, L. P. Le, G. M.
Luke, W. D. Wu, Y. J. Uemura, M. Takano, H. Dabkowska, and M. J. P.
Gingras, Phys. Rev. B \textbf{53}, 6451 (1996).

\bibitem[37]{SwFeJaro37} K. Matan, D. Grohol, D. G. Nocera, T. Yildirim,
A. B. Harris, S. H. Lee, S. E. Nagler, and Y. S. Lee, Phys. Rev. Lett.
\textbf{96}, 247201 (2006).

\bibitem[38]{CrjaroESR38} S. Okubo, R. Nakata, S. Ikeda, N. Takahashi,
T. Sakurai, W. -M. Zhang, H. Ohta, T. Shimokawa, T. Sakai, K. Okuta,
S. Hara, and H. Sato, J. Phys. Soc. Jpn. \textbf{86}, 024703 (2017).

\bibitem[39]{Spectrometer39} Y. Ihara, K. Hayashi, T. Kanda, K. Matsui,
K. Kindo, and Y. Kohama, RSI \textbf{92}, 114709 (2021).

\bibitem[40]{ref40} J. P. Perdew, K. Burke, and M. Ernzerhof, Phys.
Rev. Lett. \textbf{77}, 3865 (1996)

\bibitem[41]{ref41} G. Kresse and J. Hafner, Phys. Rev. B \textbf{47},
558 (1993).

\bibitem[42]{ref42} G. Kresse and J. Furthmüller, Phys. Rev. B \textbf{54},
11169 (1996).

\bibitem[43]{ref43} P. E. Blöchl, Phys. Rev. B \textbf{50}, 17953
(1994).

\bibitem[44]{ref44} G. Kresse and D. Joubert, Phys. Rev. B \textbf{59},
1758 (1999).

\bibitem[45]{ref45} S. L. Dudarev, G. A. Botton, S. Y. Savrasov,
C. J. Humphreys, and A. P. Sutton, Phys. Rev. B \textbf{57}, 1505
(1998).

\bibitem[46]{FP-46} J. Rodríguez-Carvajal, Physica B \textbf{192},
55 (1993).

\bibitem[47]{Shannon47} R. D. Shannon, Acta Crystallographica. \textbf{32},
751 (1976).

\bibitem[48]{Vesta48} K . Momma, and F. Izumi, J. Appl. Crystallogr.
\textbf{44}, 1272, (2011).

\bibitem[49]{Curie-Fit49} Depending on the chosen fitting range to
fit the Curie-Weiss law, a variation in the values of $\varTheta$
and $\mu_{eff}$ was observed as a result of varying temperature independent
susceptibility ($\chi_{0}$).

\bibitem[50]{Cpscaling50} M. Bouvier, P. Lethuillier, and D. Schmitt,
Phys. Rev. B\textbf{ 43}, 13137 (1991).

\bibitem[51]{FMCp42} C. Kittel, Introduction to Solid State Physics,
John Wiley \& Sons, Seventh edition, Singapore (1996).

\bibitem[52]{Mermin Wagner52} N. D. Mermin and H. Wagner, Phys. Rev.
Lett. \textbf{17}, 1133 (1966).

\bibitem[53]{Ag3LiMn2O643} R. Kumar, Tusharkanti Dey, P. M. Ette,
K. Ramesha, Atasi Chakraborty, I. Dasgupta, R. Eremina, Sándor Tóth,
A. Shahee, S. Kundu, M. Prinz-Zwick, A. A. Gippius, H. A. Krug von
Nidda, N. Büttgen, P. Gegenwart, and A. V. Mahajan, Phys. Rev. B \textbf{99},
144429 (2019). 

\bibitem[54]{dm_ref} H. J. Xiang, E. J. Kan, S.-H. Wei, M.-H. Whangbo,
and X. G. Gong, Phys. Rev. B \textbf{84}, 224429 (2011). 

\bibitem[55]{Kagome order 55} L. Messio, C. Lhuillier, and G. Misguich,
Phys. Rev. B. \textbf{83}, 184401 (2011).

\bibitem[56]{NMR Area44} Below a field of 3~T, $^{31}$P NMR measurements
attempt remained unsuccessful because of the low-frequency ringing
issue. We did see two distinct NMR lines at room temperature while
measuring the $^{31}$P - NMR spectrum at a magnetic field of strength
\textasciitilde{} 13~T.

\bibitem[57]{Moriya45} T. Moriya, Phys. Rev. \textbf{101}, 1435 (1956).

\bibitem[58]{Orbach 57} R. Orbach, Proc. R. Soc. Lond. A\textbf{
264}, 458 (1961).

\bibitem[59]{Heat capacity46} M. Evangelisti, F. Luis, L. J. de Jongh,
and M. Affronte, J. Mater. Chem. \textbf{16}, 2534 (2006). 

\bibitem[60]{Schottky 60} At high temperatures, there is a large
uncertainity in determining the magnetic specific heat as the specific
heat is mainly dominated by the lattice contribution, thus the Schottky
fit deviates from the experimental data. In order to minimize the
error in deterimining the Schottky gap, especially at the higher temperature
side, we also estimated the gap using another nonmagnetic analogue,
Li$_{9}$Ga$_{3}$(P$_{2}$O$_{7}$)$_{3}$(PO$_{4}$)$_{2}$ (LGPO),
and the error in gap size was chosen by comparing the absolute error
resulted in $\frac{\Delta_{S}}{k_{B}}$ by using LAPO \& LGPO. We
then used this result to compare with the NMR gap size.
\end{thebibliography}
\end{document}